\documentclass[12pt, onecolumn, table]{IEEEtran}
\usepackage{subfigure}
\usepackage{epsfig,graphicx,cite,amssymb,amsmath}
\usepackage{fancybox}
\usepackage{graphicx}
\usepackage{epstopdf}
\usepackage{color}
\usepackage{algorithm,algorithmic}
\usepackage{url}
\usepackage[dvipsnames,svgnames,x11names]{xcolor}
\begin{document}

%
\title{Comparative evaluation of state-of-the-art algorithms for SSVEP-based BCIs}

%
%
%

 \author{Vangelis~P.~Oikonomou,~Georgios~Liaros,~Kostantinos~Georgiadis,~Elisavet~Chatzilari,~Katerina~Adam,~Spiros~Nikolopoulos~and~Ioannis~Kompatsiaris}

%
%

\markboth{Technical Report - arXiv.org  January~2016}%
{Shell \MakeLowercase{\textit{et al.}}: Bare Demo of IEEEtran.cls for Journals}
%



\maketitle
\begin{abstract}
Brain-computer interfaces (BCIs) have been gaining momentum in making human-computer interaction more natural, 
especially for people with neuro-muscular disabilities. Among the existing solutions the systems relying on
electroencephalograms (EEG) occupy the most prominent place due to their non-invasiveness. 
However, the process of translating EEG signals into computer commands is far from trivial, since it requires 
the optimization of many different parameters that need to be tuned jointly. In this report, we focus 
on the category of EEG-based BCIs that rely on Steady-State-Visual-Evoked Potentials (SSVEPs) and perform 
a comparative evaluation of the most promising algorithms existing in the literature. More specifically,
we define a set of algorithms for each of the various different parameters composing a BCI system (i.e. 
filtering, artifact removal, feature extraction, feature selection and classification) and study each 
parameter independently by keeping all other parameters fixed. The results obtained from this evaluation  
process are provided together with a dataset consisting of the 256-channel, EEG signals of 11 subjects, 
as well as a processing toolbox for reproducing the results and supporting further experimentation. 
In this way, we manage to make available for the community a state-of-the-art baseline for SSVEP-based 
BCIs that can be used as a basis for introducing novel methods and approaches.  

\end{abstract}


%


\section{Introduction}\label{section:introduction}


An Electroencephalogram (EEG) can be roughly defined as the signal which corresponds to the mean electrical activity of the brain cells in different locations of the head. It can be acquired using either intracranial electrodes inside the brain or scalp electrodes on the surface of the head. 
Some of these electrodes are typically used as references and are either located on the scalp or on other parts of the body, e.g., the ear lobes. To ensure reproducibility among studies an international system for electrode placement, the 10-20 international system, has been defined. In this system the electrodes' locations are related to specific brain areas. For example, electrodes O1, O2 and Oz are above the visual cortex. Each EEG signal can therefore be correlated to an underlying brain area. Of course this is only a broad approximation that depends on the accuracy of the electrodes' placement.

The EEG has been found to be a valuable tool in the diagnosis of numerous brain disorders. 
Nowadays, the EEG recording is a routine clinical procedure and is widely regarded as the 
physiological ``gold standard'' to monitor and quantify electric brain activity. The electric 
activity of the brain is usually divided into three categories: 1) bioelectric events produced 
by single neurons, 2) spontaneous activity, and 3) evoked potentials. 
EEG spontaneous activity is measured on the scalp or on the brain. Clinically meaningful frequencies lie between 0.1Hz and 100Hz. 
Event-related potentials (ERPs) are the changes of spontaneous EEG activity related to a specific 
event. ERPs triggered by specific stimuli, visual (VEP), auditory (AEP), or somatosensory (SEP), 
are called evoked potentials (EP). It is assumed that ERPs are generated by activation of specific 
neural populations, time-locked to the stimulus, or that they occur as the result of reorganization 
of ongoing EEG activity. The basic problem in analysis of ERPs is their detection within the 
larger EEG activity since ERP amplitudes are an order of magnitude smaller than that 
of the rest EEG components.

When the stimulation frequency is at low rate (\textless 4Hz) the potentials are called transient VEPs while 
stimulation on higher rate (\textgreater 6Hz) produces Steady State VEPs (SSVEPs) \cite{acns:2008}.
More specifically, if a series of identical stimuli are presented at high frequency (e.g., 8 Hz), 
the system will stop producing transient responses and enter into a steady state, in which the visual 
system resonates at the stimulus frequency \cite{LuckERPbook}. In other words, when the human eye is excited by a visual stimulus, the brain generates electrical activity at the same (or multiples of) frequency of the visual stimulus. Besides the significance of SSVEPs in clinical studies, their employment as a basic building block of Brain Computer Interfaces (BCIs) make them a very important tool.


BCI gives us the ability to communicate with the external world without using peripheral nerves and muscles. A BCI system can be characterized in a number of ways based on the different modalities of physiological measurement (electroencephalography (EEG) \cite{Guger:2001,Pfurtscheller:2006}, electrocorticography (ECoG) \cite{Hill:2006} magneto-encephalography (MEG), magnetic resonance imaging (MRI) \cite{Weiskopf:2004,Yoo:2004}, near-infrared spectroscopy (fNIRS) \cite{Matthews:2008}, mental activation strategies (dependent versus independent) and the degree of invasiveness. From the above modalities, the EEG signal is the most frequently used because of its noninvasiveness, its high time resolution, ease of acquisition, and cost effectiveness compared to other brain activity monitoring modalities.

\begin{figure}[!t]
\begin{center}
\includegraphics[width=8.0cm]{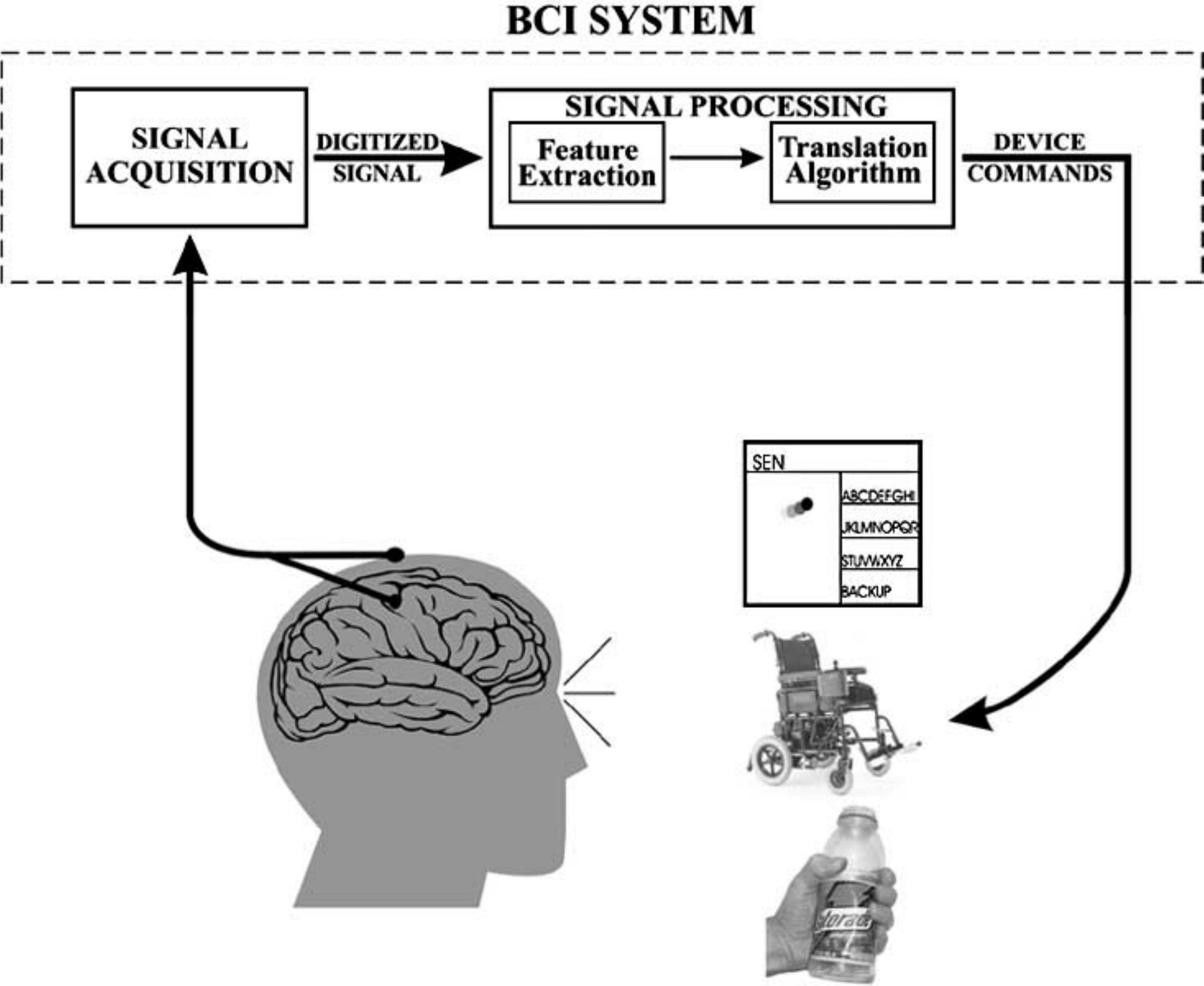}
\caption{A general description of a BCI system (reprinted from \cite{Schalk:2004})}
\label{fig:bci}
\end{center}
\end{figure}

A BCI system translates the recorded electric brain activity to output commands. To achieve that, a number of steps are performed, as indicated in Figure~\ref{fig:bci}. The input of a BCI system is the electrophysiological brain activity, while the output is the device commands. The brain activity is recorded through the use of an EEG system. After that, the analysis of EEG signals is performed in order to extract  the intended commands of the user. Different electrophysiological sources for BCI control include event related synchronization/desynchronization (ERS/ERD), VEP, SSVEP, slow cortical potentials (SCP), P300 evoked potentials and $\mu$ and $\beta$ rhythms.


\begin{figure}
\begin{center}
\includegraphics[width=14.0cm]{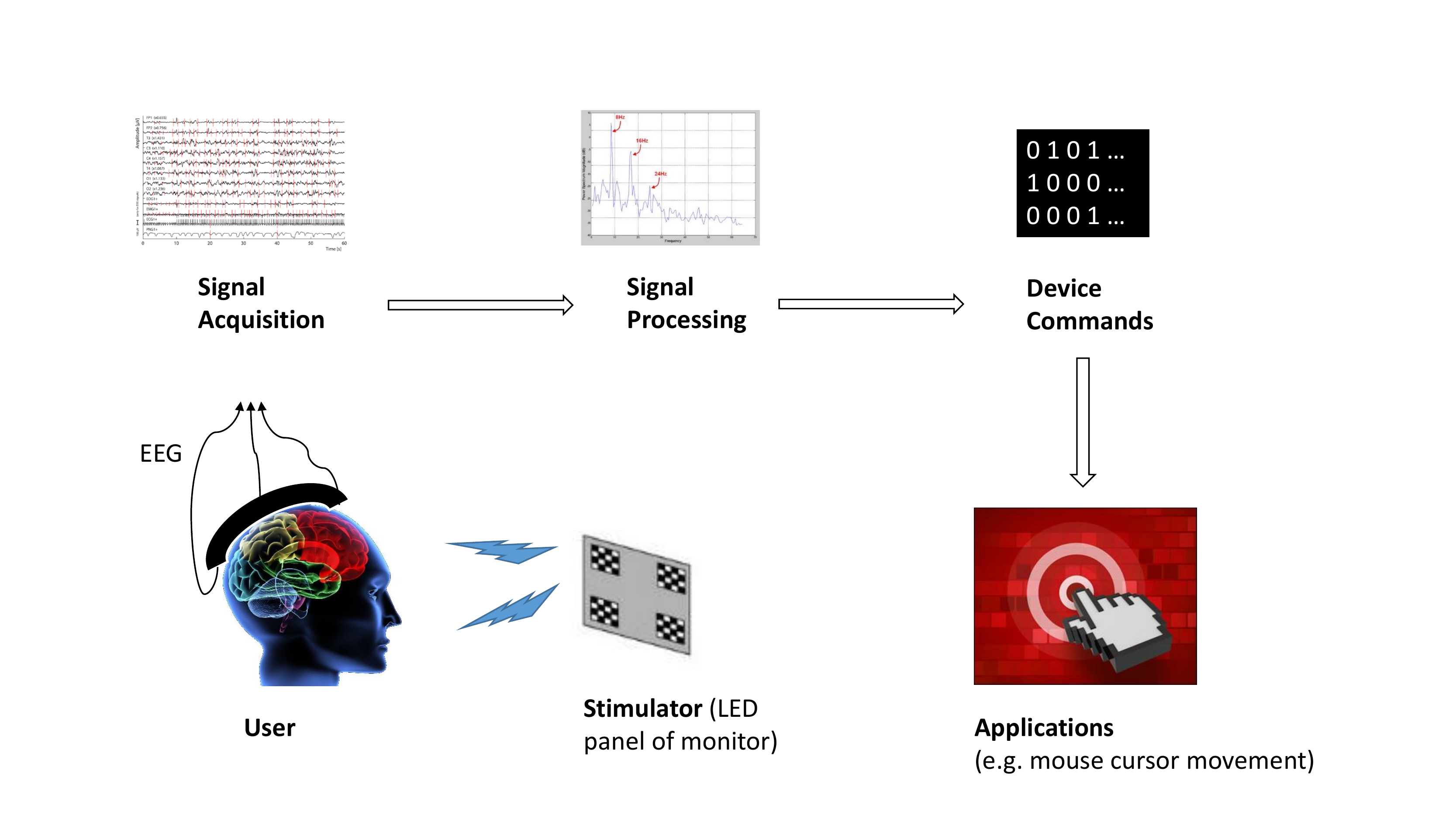}
\caption{Basic parts of a SSVEP-based BCI system}
\label{fig:ssvep_bci_parts}
\end{center}
\end{figure}

An SSVEP-based BCI (Figure~\ref{fig:ssvep_bci_parts}) enables the user to select 
among several commands that depend on 
the application, e.g. directing a cursor on a computer screen. Each command is associated with a repetitive visual stimulus that has distinctive properties (e.g., frequency). The stimuli are simultaneously presented to the user who selects a command by focusing his/her attention on the corresponding stimulus. When the user focuses his/her attention on the stimulus, a SSVEP is produced that can be observed in the oscillatory components of the user's EEG signal, especially in the signals generated from the primary visual cortex. 
In these components we can observe the frequency of the stimulus, as well as its harmonics. SSVEPs can be produced by repetitively applying visual stimuli to the user with frequencies higher to 6Hz. Compared to other brain signals (e.g. P300, sensorimotor rhythms, etc)  used for BCI 
approaches, SSVEP-based BCI systems have the advantage of achieving higher accuracy and higher information transfer rate (ITR). In addition, short/no training time and fewer EEG channels are required \cite{Amiri:2013}.

A SSVEP BCI system contains the following modules: a) Stimulator module: is a LED panel or a monitor responsible to produce the visual stimuli at a specific frequency; b) Signal acquisition module: is responsible to acquire the EEG signals during the system operation; c) Signal processing module: is responsible for the analysis of EEG signals and the translation/transformation of them into meaningful ``codewords''; and d) Device commands module: is appointed with the task to translate the ``codewords'' into interface commands according to the application setup.

\begin{figure}
\begin{center}
\includegraphics[width=12.0cm]{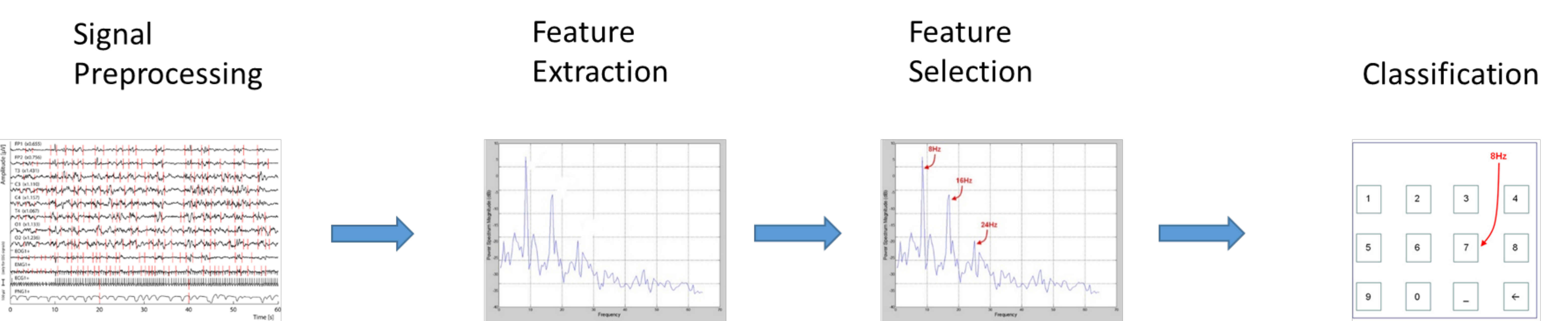}
\caption{Basic parts of the Signal Processing module in a SSVEP-based BCI system}
\label{fig:SigModule}
\end{center}
\end{figure}

Out of the aforementioned modules our interest lies on the Signal Processing module (Figure~\ref{fig:SigModule}). In the typical case, the signal processing module consists of four submodules: a) preprocessing, b) feature extraction, c) feature selection and 
d) classification. The first three submodules have the goal to make the data suitable for the classification process, which will gives us the appropriate ``codewords''. Our goal in this paper is to thoroughly examine and compare the algorithms and methods that are most widely used to implement the functionality of the aforementioned submodules, so as to obtain a state-of-the-art baseline for SSVEP-based BCI systems. Towards reaching this goal we define a set of algorithms for each submodule and adopt an empirical approach where the best algorithm for each submodule is studied independently by keeping all other submodules fixed. This allows us to obtain a close to optimal configuration for the algorithms composing the Signal Processing module without undertaking the tedious process of testing all possible combinations.  

The rest of the paper is organized as follows. Section~\ref{section:relatedwork} reviews the related literature, while Section~\ref{section:problemformulation} formulates the problem and introduces the basic notations used throughout this document. In Section~\ref{section:algorihmsandmethods} we briefly present the algorithms and methods that have been considered in our comparative study. Subsequently, in Section~\ref{section:dataacquisitionprotocol} we provide details about the protocol that has been used to obtain our experimental dataset and the processing toolbox that has been developed to obtain our results. Section~\ref{section:experimentalstudy} presents our experimental study and Section~\ref{section:conclusions} concludes this document by summarizing the most important of our findings and suggesting avenues for future research. 


\section{Related Work}\label{section:relatedwork}

The study of SSVEP-based BCIs has attracted a lot of attention in what refers to the use of algorithms and methods for maximizing the classification accuracy and improving the information transfer rate. The novelties that have been introduced in the literature cover the full spectrum of the Signal Processing module, ranging from signal filtering and artifact removal all the way to feature extraction and classification. In the following, we review the related literature along these lines.

Many methods have been applied in the preprocessing part of a SSVEP-BCI system.
The most common of them is the filtering, and most specifically the bandpass filtering.
Various filters have been used at this point of analysis procedure depending of the particular
needs of each SSVEP-BCI system. For example in \cite{Lin2007} a bandpass IIR filter 
from 22-48Hz is used to keep the desired
parts of the EEG signal. A similar IIR filter is adopted in \cite{Carvalho:2015}. 
In another work \cite{Bakardjian:2010} FIR filters are adopted to implement a filterbank. 
In addition, the filterbank approach is preferred to divide the EEG signal into bands
\cite{Chen:2015b,Martinez:2007} for further processing and analysis of EEG data. 
Besides classical time domain filtering approaches spatial filters are also used. 
More specifically, the Common Averaging Re-referencing (CAR) spatial filtering method is 
used in \cite{Carvalho:2015} to spatially filter the multichannel EEG signals and remove 
unwanted components such as eye blinks. 
Furthermore, the Minimum Energy algorithm is used in \cite{Friman:2007} to reduce 
the signal - to - noise ratio between specific EEG channels. In the same spirit the 
method of Common Spatial Pattern (CSP) is adopted in \cite{Molina:2011,Parini:2009}.
Finally, the AMUSE method in \cite{Martinez:2007} and the Independent 
Component Analysis (ICA) in \cite{Wang:2006} are used in order to remove the noise from 
multichannel EEG signals.

The notion of frequency plays a central role in SSVEP BCI systems. It is typically used to generate characteristic features in schemes that rely on classification. Thus, we must deal with this issue with great caution since it affects (and it is affected by) various factors such as the experimental stimulus presentation setup, the method that we use to 
estimate the resulting features (spectral analysis) and the classifier that it is used to assign a frequency into a class.
Spectral analysis methods are used to estimate/extract the frequency of SSVEP EEG signals. 
More specifically, the periodogram approach is used to estimate the spectral characteristics of EEG signal 
in \cite{Diez:2011,Vilic:2013,Muller:2005,Wang:2006}. Also, a more advanced method, 
the Welch algorithm, is used in \cite{Carvalho:2015}. In addition features from 
time - frequency domain, using the spectrogram, are studied in \cite{Carvalho:2015}. 
Another characteristic related to the 
frequency, and depending of the stimulus design, is the phase. 
This characteristic is exploited in \cite{Jia:2011}. Finally, time domain features, such as weighted combinations
of EEG samples, are used in \cite{Martinez:2007,Guger:2012}. Moreover, after extracting the features, a feature processing step can be introduced to further enhance the discriminative abilities of features. In this step, a selection or combination of features is adopted. More specifically, in \cite{Resalat:2013,Vilic:2013,Wang:2006} spectral features are combined 
empirically before feeding them into the classifier. A more advanced approach of selecting features is proposed in \cite{Carvalho:2015}, where an incremental wrapper is used at this stage of processing. 

Finally, the decision step in SSVEP BCI system is performed by applying a classification procedure.
More specifically, in \cite{Carvalho:2015} classifiers such as 
the Support Vector Machine (SVM), the Linear Discriminant Analysis (LDA) and 
Extreme Learning Machines (ELM) are used. SVM and LDA are the most popular classifiers
among SSVEP community and have been used in numerous works 
\cite{Carvalho:2015,Resalat:2013,Guger:2012,Rajesh:2014}. 
Furthermore, the adaptive network based 
fuzzy inference system classifier is used in \cite{Martinez:2007}. Also, neural 
networks (NN) have been used in \cite{Rajesh:2014}.
In \cite{Friman:2007} a statistic test is utilized in order to 
perform the decision, while in \cite{Vilic:2013} a set of rules is applied on spectral features.
In addition at this stage of procedure the Canocical Correlation Analysis is used. 
More specifically, 
in \cite{Lin2007} correlation indexes using the Canonical Correlation Analysis (CCA) have been
produced in order to perform the decision. Furthermore,
in \cite{Zhang:2013,Chen:2015b} more advanced usage of CCA is adopted in order to produce similar 
indexes. Finally, a similar approach is proposed in \cite{Zhang:2012b} where 
a sparse regression model was fitted to the EEG data and 
the regression coefficients are utilized for the decision.

The existence of various options for the implementation of each submodule has motivated a non-trivial number of comparative studies for BCI systems that have been reported in the literature.
In \cite{Rajesh:2014} a comparison study was presented with respect 
to the classification technique. However, the comparison was limited between SVM and NN.
Furthermore, in the feature extraction stage only features produced by FFT are used.
In \cite{Carvalho:2015} a more exhaustive comparative study has been presented. 
More specifically, in the feature extraction stage three different data sets 
have been produced based on spectral analysis, filterbank theory and 
time - frequency domain. In addition in the feature selection stage, three
feature selection approaches are used, two filters, the Pearson's filter and
the Davies Bouldin (DB) index, and one wrapper algorithm. A more thorough comparative 
study is presented in \cite{Lotte:2007} with respect to BCI systems. In this work the study was concentrated around numerous classification algorithms and the application of them in various BCI systems. 

Motivated by the same objective, in this work we perform a systematic comparison of the algorithms and methods that have been reported in the literature for SSVEP-based BCIs. Our contribution, compared to existing studies, can be summarized in the following: a) our emphasis in evaluating a system that doesn't foresee any subject-specific training prior to operation, which resulted in the adoption of the leave-one-subject-out evaluation protocol described in Section~\ref{subsection:evaluationprotocol}; b) the employment of an empirical approach for multiple parameter selection that allowed us to obtain a close to optimal configuration without having to exhaustively evaluate all possible algorithmic combinations; c) the availability of 256 channels for the EEG signals (\~40 for the occipital area) that allowed us to make some very interesting remarks on the effectiveness of different electrodes, which would have been difficult to derive with fewer channels. Finally, it is important to note that this report comes along with a dataset and a processing toolbox that have been made public for reproducing the reported results and supporting further experimentation.

\section{Problem Formulation}\label{section:problemformulation}

Let's assume that a SSVEP experiment is run with $N_s$ subjects, each of whom is presented with $N_t$ visual stimuli (a colored light flickering at different frequencies $F^{freq} = \{freq_1,freq_2,...,freq_{N_{freq}}\}$, where $N_{freq}$ is the number of flickering frequencies) for a fixed duration. Each presentation of a visual stimulus corresponding to a frequency $freq_j$ is called a trial $t_i, i=1,...,N_t$. During each trial $t_i$ we capture the EEG signal $eeg(t_i,s_k)$ of the subject $s_k$. Note that in the case of SSVEPs $freq_j$ are the labels which we want to predict (i.e. correspond to the ``codewords'' mentioned above).

After the collection of the signals, we proceed with the signal processing steps shown in Figure~\ref{fig:SigModule}. First, during the preprocessing step, we apply filtering and artifact removal to the EEG signal and we get the filtered $eeg_f(t_i,s_k)$ and the artifact-free $eeg_a(t_i,s_k)$ signals, respectively. Afterwards, each signal is typically transformed into the frequency domain that results in a set of features $eeg_w(t_i,s_k)$, Then, optionally, feature selection or dimensionality reduction can be applied to the features with the aim to increase the discrimination capacity of the resulting feature space $eeg_d(t_i,s_k)$. In the end, if we decide to employ all aforementioned processing steps each EEG signal is represented by $eeg_{f,a,w,d}(t_i,s_k)$. 

After completing the aforementioned processing steps, we have a labeled dataset consisting of pairs $\{eeg_{f,a,w,d}(t_i,s_k), freq_j\}$, for each subject $s_k$ and trial $t_i$, and its label $freq_j$. This set of labeled pairs is split into train and test set, so as to facilitate the learning and testing of a classification model. More specifically, we employ a leave-one-subject-out cross validation scheme (see also Section~\ref{subsection:evaluationprotocol}) where the labeled pairs of all subjects except $s_m$ constitute the training set and the objective is to predict the flickering frequencies of all trials undertaken by subject $s_m$ (i.e. testing set). For simplifying the notation, let us denote by $L= \{\mathbf{x}_{i}, y_{i}\}$ (with $i=\{1,\ldots,N_{L}\}$ and $y_i \in F^{freq}$) the feature vectors and associated labels that correspond to all trials expect the ones generated from subject  $s_m$, and by $U= \{\mathbf{x}_{i}, y_{i}\}$ (with $i=\{1,\ldots,N_{U}\}$ and $y_i \in F^{freq}$) the feature vectors and associated labels for the trials generated from subject  $s_m$. This means that for the training set the index of $\mathbf{x}$ runs through the trials of all subjects except $s_m$, i.e. $N_L = (N_s-1)\cdot N_t$, while for the test set the index of $\mathbf{x}$ runs through the trials generated by $s_m$, i.e. $N_U = N_t$. Thus, given the labeled training set $L$, the objective is to learn a model that will be able to estimate a score indicating whether the stimulus flickering at $freq_j$ is the source of the EEG signal represented by $\mathbf{x} \in U$ (i.e. $P(freq_j|\mathbf{x})$). Eventually, the EEG signal is classified to the frequency $\hat{freq}$ which maximizes this score: 

\begin{equation} \label{eq:ArgMaxClassification}
\hat{freq} = \arg\max\limits_{j} P(freq_j|\mathbf{x})
\end{equation}




The notations used throughout this paper can be seen in Table~\ref{tb:notation}.

\begin{table}[!ht]
\begin{center}
\caption{Notation Table}
\rowcolors{1}{Gray!25}{Gray!5}
\begin{tabular}[c]{p{5cm}p{8cm}} \hline
Notation & Meaning \\ \hline\hline
$s_i, i=1,...,N_s$ & The set of $N_s$ subjects \\
$t_i, i=1,...,N_t$ & The set of $N_t$ trials for each subject \\
$F^{freq} = \{freq_1,freq_2,...,freq_{N_{freq}}\}$ & The set of $N_{freq}$ frequencies of the visual stimuli \\
$eeg(t_i,s_k)$ & The EEG signal of subject $s_k$ for the trial $t_i$ \\
$eeg_f(t_i,s_k)$ & The filtered EEG signal of subject $s_k$ for the trial $t_i$ \\
$eeg_a(t_i,s_k)$ & The artifact-free EEG signal of subject $s_k$ for the trial $t_i$ (after artifact removal) \\
$eeg_w(t_i,s_k)$ & The transformation of the EEG signal into the frequency domain \\
$eeg_d(t_i,s_k)$ & The feature representation of the EEG signal after feature selection or dimensionality reduction \\
$N_{L} = \left\| {L} \right\| = (N_s-1)\cdot N_t$ & The total number of instances (trials) of all subjects except $s_m$\\
$L= \{\mathbf{X},Y\} = \{\mathbf{x_i},y_i\}, i=\{1,\ldots,N_{L}\}$ and $y_i \in F^{freq}$ & The train set consisting of the final feature representations of the EEG signals for all subjects except $s_m$\\
$N_{U} = \left\| {U} \right\| = N_t$ & The total number of instances (trials) of the subject $s_m$\\
$U= \{\mathbf{X},Y\} = \{\mathbf{x_i},y_i\}, i=\{1,\ldots,N_{U}\}$ and $y_i \in F^{freq}$ & The test set consisting of the final feature representations of the EEG signals for subject $s_m$ \\
\end{tabular}
\label{tb:notation}
\end{center}
\end{table}

\section{Algorithms and Methods}\label{section:algorihmsandmethods}

In our effort to achieve a state-of-the-art baseline for SSVEP-based BCIs, we have examined a number of algorithms and methods that are most widely used in the respective area. In this section we specify with more detail the different algorithms and methods that have been used to implement the different submodules of the signal processing module (see Figure~\ref{fig:SigModule}).

\subsection{Signal pre-processing}\label{subsection:signalpreprocessing}

Inside the signal preprocessing step the data are processed in order to remove the unwanted components of the signal. More specifically, the EEG signals are likely to carry noise and artifacts. For instance, electrocardiograms (ECGs), electrooculograms (EOG), or eye blinks affect the EEG signals. In order to remove this noise and artifacts, we can rely on the fact that EEG signals contain neuronal information below 100 Hz (in many applications the information lies below 30 Hz), so any frequency component above these frequencies can be simply removed by using lowpass filters. Similarly a notch filter can be applied to cancel out the 50Hz line frequency. However, when the components generated by noise and artifacts lie at the effective range of EEG signals more sophisticated methods of pre-processing are necessary.

\subsubsection{Signal Filtering}\label{subsec:signalfiltering}

Between filters and spectral analysis a strong relation exists, since the goal of filtering is 
to reshape the spectrum of a signal in our advantage. The way that we achieve
this reshaping defines a particular filter. There is two general groups of filters, 
Finite Impulse Response (FIR) filters and Infinite Impulse Response (IIR) filters that are 
characterized by their impulse response. FIR filters refer to filters that have an impulse 
response of finite duration, while  (IIR) filters have an impulse response of infinite duration. 
Any linear system, describing the relation between the input signal $eeg$ and the output (i.e. filtered) signal $eeg_f$, is given by: 

\begin{equation}\label{EqLinearSystem}
eeg_f[n] = \sum_{k=1}^{K}a[k]eeg_f[n-k] + \sum_{m=0}^{M}b[m]eeg[n-m]
\end{equation}
where $eeg_c[n]$ is the output and $eeg[n]$ the input signal to the system at the $n$-th time point and $a[k],k=1,\cdots,K,b[m],m=1,\cdots,M$ are the coefficients of the linear system. Any IIR filter is described
by Eq. (\ref{EqLinearSystem}), while any FIR filter can be described by Eq. (\ref{EqLinearSystem}) when we set
the $a[k]$ coefficients to zero.

Both filter types can achieve similar results with respect to the filtering process. However differences between them exists. The FIR filters
are always stable and present the characteristic of linear phase \cite{OppenheimDSP:1999,ProakisDSP:1996,White:2000}. 
In contradiction, IIR filters are not always stable and present nonlinear phase
characteristics\cite{OppenheimDSP:1999,ProakisDSP:1996}. However, IIR filters require fewer coefficients than FIR filters,
which make them suitable for cases where the memory constraints are critical and some phase distortion is
tolerable\cite{OppenheimDSP:1999,ProakisDSP:1996}. 
An extensive comparison between the two types is outside the scope of this work. 
The interested reader is referred to \cite{OppenheimDSP:1999,ProakisDSP:1996,White:2000} for a more thorough discussion on this subject. 

The most typical approach to remove noise from the EEG signal is to use band-pass filters, provided that the frequencies of the noise components do not overlap with the frequencies that convey the phenomena of interest.
Band pass filters work by attenuating the frequencies of specific ranges while amplifying others. 
In our case, we mainly focus on the range of 5-48Hz, which is the range defined by the stimuli frequencies (see Section~\ref{section:dataacquisitionprotocol}) and their harmonics (up to the 4\textsuperscript{th} order). This means that we can remove a large portion of noise related with EMG (high frequency noise), EOG (low frequency noise) and electrical current (in our case 50Hz).

\subsubsection{Artifact removal}\label{subsubsec:artifactremoval}

Despite the filtering of the signal some noise may still persist, such as the EOG artifacts that may be present in the 0-10Hz frequency range, such as the artifacts generated from eye blinks. For this purpose we need to follow an additional pre-processing step that is generally addressed as artifact removal. In the following, we provide details for two of most widely used approaches for artifact removal, namely AMUSE and Independent Component Analysis. 

\paragraph{AMUSE}

Some of the existing techniques for artifact removal rely on blind source separation (BSS), with AMUSE \cite{Choi:2005} being one of the most typical representatives. The AMUSE algorithm  have been used  previously for artifact removal in SSVEP analysis by \cite{Martinez:2007} and belongs to the second-order statistics spatio-temporal decorrelation algorithms \cite{Choi:2005}. AMUSE consists of two steps and each step is based on principal component analysis (PCA). In the first step PCA is applied for whitening the data, while in the second step the singular value decomposition (SVD) is used on a time delayed covariance matrix of the pre-whitened data. 

For this section let us denote the signal that is captured from all channels of the EEG sensor during a trial as $\mathbf{r} = [\mathbf{r}(1),\ldots,\mathbf{r}(N_{channels})]$, a matrix that contains the observations from all channels over time, with $\mathbf{r}(n)$ being the vector that contains the observations from all channels in time $n$. Let us also assume that $\mathbf{R}_{\mathbf{r}}$ is the covariance matrix, $\mathbf{R}_{\mathbf{r}}=E\{\mathbf{r}(n)\mathbf{r}^T(n)\}$. The goal of AMUSE is to decompose the observations into uncorrelated sources, $\mathbf{z}(n)=\mathbf{W}\mathbf{r}(n)$; or to make sure that the observations are produced by linearly mixing uncorrelated sources, $\mathbf{r}(n)=\mathbf{A}\mathbf{z}(n)$. 
The first step of AMUSE aims to find a linear transformation for whitening the data, 
$\mathbf{z}(n)=\mathbf{Q}\mathbf{r}(n)$. The matrix $\mathbf{Q}$ that satisfies the whitening property is $\mathbf{Q}=\mathbf{R}_{\mathbf{r}}^{-\frac{1}{2}}$. Next, the SVD is 
applied on a time delayed covariance matrix of the whitened data $\mathbf{z}(n)$, $\mathbf{R}_{\mathbf{z}(n)\mathbf{z}(n-1)} = \mathbf{U}\mathbf{\Lambda}\mathbf{V}$, where $\mathbf{\Lambda}$ is the diagonal matrix with decreasing singular values and $\mathbf{U}$, $\mathbf{V}$ are the orthogonal matrices of singular vectors. In the case of AMUSE the unmixing matrix is estimated as $\mathbf{W}=\mathbf{U}^T\mathbf{Q}$, so the estimated components are given by:
\begin{equation}
\mathbf{\hat{Z}} = \mathbf{W}\mathbf{r}
\end{equation}

\noindent where $\mathbf{r}$ is the matrix that contains the observations from all channels over time. A useful characteristic of AMUSE is that it provides us with a ranking of components, due to the application of SVD, and this ranking is used to remove the unwanted components from the EEG signals. Based on the results reported in the literature\cite{Martinez:2007}, eye blinks and other kinds of noise are typically considered to lie in the few first and last components generated by AMUSE.  Subsequently, the remaining components can be projected back to the original space using the pseudo - inverse of the unmixing matrix, aiming to yield and artifact-free version of the data:

\begin{equation}
\mathbf{\hat{r}} = \mathbf{W}^{+}\mathbf{\hat{Z}}
\end{equation}

\noindent Finally, the artifact free signal $eeg_a$ is the row of the matrix $\mathbf{\hat{r}}$ corresponding to the channel we want to use.

\paragraph{Independent Component Analysis:} Another category of methods for artifact removal relies on the use of Independent Component Analysis (ICA). Consider the matrix $\mathbf{r}$ of observations denoted in the previous section.
In ICA we assume that each vector $\mathbf{r}(n)$ is a linear mixture of $K$ unknown sources $\mathbf{z}(n) = \{{z}_1(n),\ldots,{z}_K(n)\}$:
\begin{equation}
\mathbf{r}(n) = \mathbf{A}\mathbf{z}(n) \nonumber
\end{equation}
where the matrix of mixing coefficients $\mathbf{A}$ is unknown. The goal in ICA is to find the sources $z_i(n)$, or to find the inverse of the matrix $\mathbf{A}$. The sources are independently distributed with marginal distributions $p(\mathbf{z}_n)=p_i(z(n)^{(i)})$.
The algorithm to find the independent components has three steps: a) Calculate an estimation sources through the mapping: $\mathbf{a} = \mathbf{Wr}$; b) Calculate a nonlinear mappping of the estimated sources $b_i=\phi(a_i)$. A popular choice is to use the $tanh$ function for the function $\phi$; c) adjust the matrix $\mathbf{W}$ through $\Delta \mathbf{W} \propto [\mathbf{W}^{-1}]^{T} + \mathbf{b}\mathbf{r}^T$. The above exposition of the ICA is based on the ML principle,
however similar algorithms for ICA can be obtained by adopting
other criteria for the independence. A useful introduction in ICA
is presented \cite{Hyvarinen:2000}, where a fast algorithm to
perform ICA is also given. 

From the perspective of biomedical signal processing ICA has found many applications, with the study of brain dynamics through EEG signals being among them \cite{Jung:2001a,Jung:2001b}. The general application of ICA in EEG data analysis is performed in three steps. First, the EEG data are decomposed in
independent components, then by visual inspection some of these
components are removed since they are considered to contain artifacts (for example eyes
blink), and finally the artifact-free EEG signals are obtained by
mixing and projecting back onto the original channels the selected 
non-artifactual ICA components.

Both ICA and AMUSE are used for blind source separation. However, differences
between the two approaches exist\cite{Choi:2005}. For example, ICA is based 
on high order statistics while AMUSE belongs to methods that use second order statistics. Also, AMUSE provides us automatically an ordering of components due to the application of SVD (singular value decomposition). This ordering of components can be particularly useful in devising a general rule about which components to remove, without having to inspect the EEG data every time.

\subsection{Feature Extraction}\label{subsection:featureextraction}

After signal pre-processing, the feature extraction step takes place. A feature is an alternative representation of a signal, and in most cases of lower dimensionality. Feature extraction is very important in any pattern 
recognition system, since this step determines various descriptors of the signal. Also, the choice of features has an important influence on the accuracy and the computational cost of the classification process. There are two well-known categories for extracting features: a) Nontransformed features (moments, power, amplitude information, energy, etc.) and b) Transformed features (frequency and amplitude spectra, subspace transformation methods, etc.). Transformed features form the most important category for biomedical signal processing and feature extraction. The basic idea employed in transformed features is to 
find such a transformation (e.g. Fourier Transform) of the original data that best represent the relevant information for the subsequent pattern recognition task. Feature extraction is highly domain-specific and whenever a pattern recognition task is applied to a relatively new area, a key element is the development of new features and feature extraction methods.

A number of approaches can be applied in order to extract useful features for the subsequent analysis of the data. As reported in previous sections, a useful characteristic of SSVEP-based signals is the synchronization of 
the brain to the frequency of the stimulus. To exploit this characteristic the frequencies 
contents of EEG signals must be analyzed. The study of frequencies of a signal falls into the 
concept of spectral analysis. Spectral analysis is the process of estimating how the total 
power of a finite length signal is distributed over frequency. A large number of methods 
for estimating the spectrum (or the Power Spectral Density - PSD) of a signal can be found 
in the literature. In this report we confine our study to well-known PSD methods that 
have been applied widely for SSVEP analysis. The objective of this step is, given the EEG signal $eeg[n]$, to extract a representation for it $eeg_w$ using a transformation function $P^*(f)$, where $*$ takes various values depending on the type of transformation (e.g. $eeg_w = P^{AR}(f)$ is the AR spectrum).

\subsubsection{Periodogram}
The first and most simple method for the estimation of a PSD is the
periodogram\cite{OppenheimDSP:1999,ProakisDSP:1996}. The periodogram 
is simply the Discrete Fourier Transform (DFT) of the signal.
More specifically, let’s assume the discrete signal 
${eeg[n],n=1,…,N}$ then the periodogram of $eeg[n]$ is defined as:
\begin{equation} 
P^{Per}(f)= \frac{1}{N} \Big| \sum_{n=0}^{N-1} eeg[n]e^{-j2 \pi fn}\Big|^2=\frac{1}{N}|EEG(f)|^2
\end{equation}
where $EEG(f)$ is the discrete Fourier Transform of the sequence $eeg[n]$. The periodogram provides 
some computational advantages by using the Fast Fourier Transform (FFT) algorithm to calculate 
the Fourier Transform of $eeg[n]$. However, from a statistical point of view, the periodogram is 
an unbiased and inconsistent estimator. This means that the periodogram does not converge to 
the true spectral density. Also, this estimator presents problems, related to spectral 
leakage and frequency resolution, due to the finite duration of $eeg[n]$.

\subsubsection{Welch Spectrum}
To reduce the above effects various non parametric methods for the estimation of PSD were proposed. 
These approaches use the periodogram as a basic component of the method and at the end provide a 
modified version of it. A well-known non parametric method for PSD estimation is the Welch's method\cite{OppenheimDSP:1999,ProakisDSP:1996}. 
This method consists of three basic steps:
\begin{enumerate}
\item First, divide the original $N$ – length sequence into $K$ segments (possibly overlapped) 
of equal lengths $M$.
\begin{align}
eeg_{i}[m]=eeg[m + iD], i=0,\cdots,K-1, m=0,\cdots,M-1
\end{align}
\item Apply a window to each data segment and then calculate the periodogram on the 
windowed segments (modified periodograms).
\begin{align}
P_{i}(f)= \frac{1}{NU} \Big| \sum_{n=0}^{N-1} w[m]\cdot eeg_{i}[m]e^{-j2 \pi fm}\Big|^2, i=0,\cdots,L-1
\end{align}
where $U=\frac{1}{M}\sum_{m=0}^{M-1} w[m]^2$ is a normalization constant with 
respect to the window function.
\item Average the modified periodograms from the K segments in order to obtain an estimator 
of the spectral density.
\begin{align}
P^{W}(f) = \frac{1}{K} \sum_{i=0}^{K-1} P_{i}(f)
\end{align}
\end{enumerate}
The Welch estimator is a consistent estimator, but continues to present the problems of 
frequency resolution and spectral leakage, although the effects of the above problems 
are reduced compared to the periodogram.

\subsubsection{Goertzel algorithm}

The basic component of the above spectrum estimation methods is the Fast Fourier Transform (FFT), 
which is used for the estimation of Discrete Fourier Transform (DFT) coefficients. However, 
in the case that we want to estimate $K$ DFT coefficients from the total $N$ when $K<<N$, an alternative
approach exists. This approach is called the Goertzel algorithm \cite{OppenheimDSP:1999,ProakisDSP:1996}. The basic idea of this algorithm
is that the computation of DFT coefficients can be obtained by a linear filtering operation.
When $K<\log_{2}N$ then this algorithm is more efficient than FFT. In our study we used the Goertzel
algorithm when a subset of frequencies is needed to be calculated.
In the case of Goertzel algorithm the spectrum values are obtained by applying the squaring operation on
the DFT coefficients. More specifically, let $EEG[f],f=1,\cdots,K$ be the DFT coefficients from 
Goertzel algorithm, then the spectrum values are obtained as: $P^{G}(f)=|EEG(f)|^2$.

\subsubsection{Yule - AR Spectrum}

The non-parametric methods, such as the Welch method, are simple, easy to implemented 
and can be calculated very fast using the FFT algorithm. However, they require large data 
records to achieve the desired frequency resolution, and, they suffer from spectral leakage 
effects due to finite duration of the data. To solve these problems parametric methods have 
been adopted for spectral estimation. The parametric methods avoid the leakage effect while 
generally provide better frequency resolution than non-parametric methods. The general idea 
is to assume a model that generates the data and then, using this model, provide an estimation of spectral density. A well – known parametric method is the Yule – Walker AR 
(PYULEAR) method\cite{stoica2005spectral}. This method assumes an autoregressive (AR) model of order $p$ for the 
generation of the sequence $eeg[n]$. Then, using the sequence $eeg[n]$, it estimates the model 
parameters. After that, the PSD is estimated according to predetermined equations. A limitation 
of the above method is how to determine the model's order $p$.
The Yule-AR spectrum is given by:
\begin{align}
P^{AR}(f) = \frac{\sigma_{e}^{2}}{|1+\sum_{k=1}^{p}a[k]e^{-j2 \pi fk}|^2 }
\end{align}
where $a[k],k=1,\cdots,p$ are the estimated AR coefficients and $\sigma_{e}^{2}$ is the estimated minimum prediction error.

\subsubsection{Short Time Fourier Transform}

All the above methods make the assumption that the data are generated from a stationary process, 
i.e. the frequency content of the sequence $eeg[n]$ does not change with the time. However, 
this assumption may not be hold always and the sequence $eeg[n]$ may present non–stationarities. 
To study the non–stationarities of a sequence the notion of time varying spectrum has been 
introduced. A first approach to obtain a time varying spectrum is the Short Time Fourier Transform 
(STFT)\cite{MallatWTS:2008}. The general idea of this approach is to divide the original sequence into segments and 
calculate the Fourier transform in each segment. Then we can plot the spectrum of each segment 
to observe the changes of frequency over time.

The Short Time Fourier Transform is given by:
\begin{align}
S[m,f] = \sum_{n=1}^{N} eeg[n]w[n-m] e^{-j2 \pi fn}  
\end{align}
where $ m=\{1,2,\cdots, N\}, f=\{0,\frac{1}{N}, \frac{2}{N},\cdots, 1\} $ and $w[\cdot]$ is a preselected
specialized window function. Finally, the spectrogram is obtained as:
\begin{align}
P^{S}[m,f] = |S[m,f]|^2.
\end{align}

\subsubsection{Discrete Wavelet Transform}

In all above presented methods, the Fourier Transform (FT) plays the most critical role in 
the method. The basic idea of FT is to represent the sequence $eeg[n]$ as a linear superposition 
of sinusoidal waves (sines and cosines). However, sinusoidal waves are not well localized in 
time and hence the FT needs many coefficients to represent localized event such as a transient 
phenomenon. An extension of the FT is the Wavelet Transform (WT) \cite{MallatWTS:2008} 
where a sequence is 
represented as a linear superposition of wavelets. Wavelets are well localized in time and 
frequency while the wavelet coefficients present sparse nature.
A wavelet basis is obtained by applying careful dilations and translations of the mother function $\psi(n)$:
\begin{align}
\Big\{ \psi_{j,k}(t) = \frac{1}{2^{j/2}} \psi\Big(\frac{t-2^{j}k}{2^j}\Big) \Big\}_{j,k\in \mathbb{Z}}
\end{align}
A wavelet atom $\psi_{j,k}(t)$ is localized around the point $ 2^{j}k$ and its support is proportional to the scale $2^j$. In the wavelet transform, the signal is represented as a linear combination of wavelets and scaling functions:
\begin{align}
\mathbf{P^{DWT}} = \mathbf{W}\mathbf{g}
\end{align}
where $\mathbf{g}$ is the vector of wavelet coefficients and $\mathbf{W}$ is a matrix containing the wavelets and scaling functions. In practice, the wavelet coefficients are obtained by using filter banks.  

\subsection{Feature selection}\label{subsection:featureselection}

Although the feature extraction methods are specifically designed to bring out those aspects of the data that are most favorable in performing the intended classification task, it is rather typical to employ an additional step that has to do with feature selection. The goal of this step is to further increase the accuracy of the classification, either by reducing the dimensionality of the feature space, and thus alleviating the curse of dimensionality, or by removing redundant information that is likely to confuse the classifier in distinguishing between the existing classes.

With respect to the former, common approaches for dimensionality reduction use techniques from linear algebra such as Principal Component Analysis (PCA),  Singular Value Decomposition (SVD) and Independent Component Analysis (ICA). With respect to the latter, the approach of Feature Subset Selection is typically followed. The general idea behind this approach is to select a subset of the available features based on some criteria, such as the entropy, the information contained in each feature, and mutual information, the amount of information shared by two different features, etc. While it seems that in the process some information will be lost, this is not the case when redundant and irrelevant features are present. The list of excluded features may contain features that do not have a significant impact on the output, or others that have a strong impact whilst they should not. In the first case the result is a more compact representation of the features with less redundancy, whereas in the second overfitting phenomena are avoided. 

In the following we examine two algorithmic categories for feature selection, the ones relying on information theory and the ones resulting from the projection of the original feature space into a set of pre-calculated components. 

\subsubsection{Shannon's Information Theory}\label{subsubsection:entropyFeatureSelection}

Based on Shannon’s Information Theory~\cite{Shannon:1948} and the measures provided below, the goal 
of entropy-based feature selection methods is to produce a Scoring Criterion (or Relevance Index) 
$\mathbf{J}$ that determines whether a feature is useful or not when used in a classifier. 
More specifically, the first measure is the entropy and measures the uncertainty present in the 
distribution of a random variable $RX$ using the probability distribution $p(rx)$ of $RX$. 
In this case, we consider the feature vectors $eeg_w$ to determine the probability density function (pdf) of $RX$.

\begin{align}
\mathbf{H}(RX) = - \sum_{rx\in{RX}} p(rx) \log p(rx)
\end{align}

Furthermore, conditional entropy, the second measure, is used to reduce uncertainty 
for $rx \in{RX}$ when $RY$ is known:
\begin{align}
\mathbf{H}(RX|RY) = - \sum_{rx\in{RX}} \sum_{ry\in{RY}} p(rx,ry) \log p(rx|ry)
\end{align}
Mutual information is the last measure and it quantifies the amount of information shared by $RX$ and $RY$. It is defined as the difference between entropy and conditional entropy:

\begin{align}
\mathbf{I}(RX;RY) = \mathbf{H}(RX) - \mathbf{H}(RX|RY) 
\end{align}

Based on these measurements, the 12 different indexes available on FEAST~\cite{Brown:2012} aim at 
reducing the redundancy and/or increasing the complementary information of the features. For this 
purpose features are sorted in a descending order based on $\mathbf{J}$ and a subset of them 
$\mathbf{S}$ (i.e. constituting $eeg_d$) is selected for the next steps of analysis. Thus, 
for each of the $RX_k$ features in the feature set ($RX$), $\mathbf{J}$ was obtained 
using the following approaches:

\begin{align}
\mathbf{Jmim}(RX_k) = \mathbf{I}(RX_k;RY)
\end{align}

Mutual Information Maximization (MIM) is the Scoring Criterion that denotes the features 
with the highest mutual information to a class $RY$, while estimating $\mathbf{J}$ 
independently for each feature.

\begin{align}
\mathbf{Jmifs}(RX_k) = \mathbf{I}(RX_k;RY) - \beta \sum_{RX_j \in{S}} I(RX_k;RX_j)
\end{align}

Mutual Information Feature Selection (MIFS) introduces a penalty factor to the 
currently selected set of features to reduce redundancy. The $\beta$ is a 
configurable parameter and when set to zero the results will be identical to MIM 
Scoring Criterion.

\begin{align}
\mathbf{Jjmi}(RX_k) =  \sum_{RX_j \in{S}} I(RX_k RX_j; RY)
\end{align}
Joint Mutual Information (JMI) criterion aims to increase the complementary information 
by including a feature to the existing set if it is complementary, while creating 
the joint random variable $RX_k RX_j$.

\begin{align}
\mathbf{Jcmi}(RX_k) = \mathbf{I}(RX_k;RY|S)
\end{align}

Conditional Mutual Information (CMI) criterion examines the information still 
shared by $RX_k$ and $RY$ after the set of all currently selected 
features ($\mathbf{S}$) is revealed.

\begin{align}
\mathbf{Jmrmr}(RX_k) = \mathbf{I}(RX_k;RY) - \frac{1}{|S|} \sum_{j \in{S}} I(RX_k;RX_j)
\end{align}

Minimum - Redundancy Maximum - Relevance (MRMR) is a criterion similar to the MIFS but 
the conditional redundancy is omitted.

\begin{align}
\mathbf{Jcmim}(RX_k) = \mathbf{I}(RX_k;RY) - \max_{RX_j \in{S}} [I(RX_k;RX_j) - I(RX_k;RX_j|RY)]
\end{align}

Conditional Mutual Information Maximization (CMIM) is a criterion that evaluates the features in 
a pairwise fashion, as opposed to the previously described criteria that evaluate each feature 
separately.

\begin{align}
\mathbf{Jicap}(RX_k) = \mathbf{I}(RX_k;RY) -\max{[0, {I(RX_k;RX_j) - I(RX_k;RX_j|RY)}]}
\end{align}

Interaction Capping (ICAP) criterion uses both mutual information and conditional mutual 
information for the evaluation of each feature.

\begin{align}
\mathbf{Jcife}(RX_k) = \mathbf{I}(RX_k;RY) - \sum_{RX_j \in{S}} I(RX_k;RX_j) + \sum_{RX_j \in{S}} I(RX_k;RX_j|RY)
\end{align}

Conditional Informax Feature Selection (CIFE) criterion, similarly to ICAP, uses both mutual 
information and conditional mutual information with the main difference being that both these 
terms are bound by $\mathbf{S}$ and not by $RX_k$.

\begin{align}
\mathbf{Jdisr}(RX_k) =  \sum_{j \in{S}} \frac{I(RX_k RX_j ; RY)}{H(RX_k RX_j RY)}
\end{align}

Double Input Symmetrical Relevance (DISR) criterion introduces a normalization term ($\mathbf{H}$) to the JMI criterion.

\begin{align}
\mathbf{Jbetagamma}(RX_k) = \mathbf{I}(RX_k;RY) + \beta I(RX_j;RX_k) +\gamma I(RX_k;RX_j|RY)
\end{align}

Beta Gamma criterion assigns weights ($\mathbf{\beta}$) and ($\mathbf{\gamma}$) to redundant mutual and 
conditional mutual information respectively. Using $\mathbf{\beta} = 0$ and $\mathbf{\gamma} = 0$ 
the results are equivalent to MIM scoring criterion

\begin{align}
\mathbf{Jcondred}(RX_k) = \mathbf{I}(RX_k;RY) + \mathbf{I}(RX_k;RX_j|RY)
\end{align}

Conditional Redundancy (CONDRED) criterion is a special case of Beta Gamma, 
where $\mathbf{\beta} = 0$ and $\mathbf{\gamma} = 1$ and eliminates the 
redundant mutual information.

\subsubsection{Data projection} \label{subsubsection:dataprojectionFeatureSelection}

Another approach for feature selection is the linear dimensionality reduction techniques like Principal Component Analysis (PCA) and Singular Value Decomposition (SVD). Both techniques aim at mapping the original data $\mathbf{X}$ to a lower dimensional space using a projection matrix. More specifically, given a data matrix PCA generates a new matrix called the principal components, with each component being the linear transformation of the data matrix. Each principal component contains the variance, with the components being sorted in a descending order. Finally, based on the intended dimensionality of the resulting feature space, we retain the corresponding number of principal components and project the original data on these components. 

In SVD, the data matrix is decomposed into 3 new matrices, $\mathbf{U}$ and $\mathbf{V}$ that are unitary transforms and $\mathbf{S}$ that is diagonal. $\mathbf{U}\times\mathbf{S}$ provide the coefficients and $\mathbf{V}$ provides the eigenvectors. The number of the selected eigenvectors defines the desired dimensionality. The data matrix $\mathbf{X}$ decomposition using SVD is defined as follows (here $\mathbf{X}$ refers to the training set as defined in Table~\ref{tb:notation}):

\begin{align}
\mathbf{X} = \mathbf{U} \mathbf{S} \mathbf{V^T} 
\end{align}

Finally, we choose to retain a number of the pre-computed data components $\hat{S}$ (i.e. eigenvectors) so as to linearly project the original feature space into a new feature space (i.e. the $eeg_d$).

\begin{align}
\mathbf{\hat{X}} = \mathbf{U} \mathbf{\hat{S}} \mathbf{V^T} 
\end{align}

Although both PCA and SVD are mostly used for reducing the dimensionality of the original space, they often increase the discrimination capacity of the data. 

\subsection{Classification}
\label{ssec:ClassificationTheory}

At the final stage of the signal processing module we encounter the classification (or pattern recognition) algorithm. Classification is the task of assigning the EEG signals into one of several predetermined categories/classes. A classifier is essentially a systematic approach for building classification models from an input data set. Examples include decision tree classifiers, rule-based classifiers, neural networks, support vector machines, and naive Bayes classifiers. Each technique employs a learning algorithm to identify a model that best fits the relationship between the features and class label of the input data. The model generated by a learning algorithm should be able to fit the input data that has been trained from, as well as to correctly predict the class labels of records that has never seen before. Therefore, a key objective of the learning algorithm is to avoid overfitting and build models with good generalization capability.

Classification can be formalized as the problem of learning a mapping function $y=f(\mathbf{x})$, which maps a feature vector $\mathbf{x}$ to a label $y$. 
Based on the nature of the label $y_i$, we can distinguish between binary classification ($y_i \in \{-1,1 \}$) and multi-class classification ($y_i \in D = \{d_1,d_2,...,d_m \}$, where $D$ is the label space and $d_j$ denotes the class). In the case of multi-class classification with $m$ classes, there are several algorithms that are inherently multi-class (e.g. decision trees, random forests, Adaboost, Naive Bayes, etc), which are usually probabilistic and graph based. On the contrary, similarity based algorithms (e.g. Support Vector Machines (SVMs), Linear Discriminant Analysis (LDA), etc) being usually binary with the objective of separating the positive from the negative class, cannot be applied directly on multi-class classification problems. A typical way to overcome this problem is to split the multi-class problem into several binary classification problems (e.g. into $m$ binary classification problems when using the one-vs-all (OVA) trick for the transformation). 

In this work we compare several popular machine learning algorithms from both categories (multi-class and binary). In the following we give a brief explanation for each of the examined algorithms.



\subsubsection{Support Vector Machines (SVMs)}
\label{sssec:SVMs}

The most popular classification algorithm is the SVMs, which aims to find the optimal hyper-plane that separates the positive class from the negative class by maximizing the margin between the two classes. This hyperplane, in its basic linear form, is represented by its normal vector $\mathbf{w}$ and a bias parameter $b$. These two terms are the parameters that are learnt during the training phase. Assuming that the data is linearly separable, there exist multiple hyper-planes that solve the classification problem. SVMs choose the one that maximizes the margin, assuming that this will generalize better to new unseen data. This hyper-plane is found by solving the following minimization problem:

\begin{eqnarray}
\label{SVM_2}
&  \min_{\mathbf{w},b} \frac{1}{2}\|\mathbf{w}\|^2 + C\displaystyle\sum_{i=1}^{N} \xi_i\nonumber \\
 \text{s.t.:} & y_i(\mathbf{w}^T\mathbf{x}+b) \ge 1-\xi_i, \xi_i \ge 0, i=1,\cdots,N
\end{eqnarray}

\noindent where $\xi_i$ are slack variables relaxing the constraints of perfect separation of the two classes and $C$ is a regularization parameter controlling the trade-off between the simplicity of the model and its ability to better separate the two classes.


In order to classify a new unseen test example $\mathbf{x_{j}} \in U$, its distance from the hyper-plane is calculated by the following equation.

\begin{equation} \label{eq:SVMdistance2Hyperplane}
f(\mathbf{x_{j}}) = \mathbf{w}^T\mathbf{x_{j}} + b
\end{equation}

It is important to note that linear SVMs are based on the assumption that the training data is linearly 
separable (i.e. there exists a set of parameters $\mathbf{w}$ that separates perfectly the two classes). 
However, this is rarely the case for real world data. For this reason the kernel trick was introduced,
which allows for the hyper-plane to take various forms (e.g. a hyper-sphere if the RBF kernel is selected). 
This was accomplished by projecting the input data into a higher dimensional space using a Kernel function 
$K(x_i,x_j)$ to measure the distance between the training instances $\mathbf{x_i}$ and $\mathbf{x_j}$. 
In this formulation, the dot product of the linear case is replaced with nonlinear kernel functions: 

\begin{equation}
\mathbf{w}^T \mathbf{x} \mapsto \sum\limits_{i=1}^{n} w_{i} y_i K(\mathbf{x}_i,\mathbf{x}))
\end{equation}

In order to transform the output of the mapping function $f(\mathbf{x_{j}})$ (i.e. the distance 
of $\mathbf{x_{j}}$ to the hyper-plane) to a probability, Platt's sigmoid is typically used~\cite{Platt99,Lin2007}. 
This allows for using the maximization equation (Eq.~\ref{eq:ArgMaxClassification}) without having to 
normalize the scores for each concept.

\subsubsection{Decision Trees}
\label{sssec:dtrees}

A decision tree is a decision support tool that uses a tree-like graph or model of 
decisions  and their possible outcomes (e.g. If the attribute $x_i \geq 0$ go to the left 
child otherwise to the left). A decision tree is a hierarchical structure consisting of two 
types of nodes; a) the internal decision nodes and b) the prediction nodes. All nodes basically 
examine whether a condition is satisfied, e.g. whether the value of a given attribute is 
higher/lower than a certain value. However, the internal decision nodes have other nodes as 
children, while prediction nodes have no children and correspond to the class labels. The 
advantages of decision trees are that they are simple to understand, implement and use, can 
learn complicated decision boundaries and support both real valued and categorical features 
(attributes). However they are prone to over-fitting. 

\subsubsection{Ensemble Learning}
\label{sssec:ensemble}

Ensemble learning, based on the assumption that multiple weak classifiers can perform better than a single but more robust classifier, trains multiple classifiers either based on the same learning algorithm (e.g. using different subsets of the training set each time) or different learning algorithms. However, usually, the term ensemble is reserved for methods that generate multiple classifiers using the same base learner (e.g. decision trees). Evaluating the prediction of an ensemble typically requires evaluating the prediction of each single weak classifier and combining them. In this way, ensembles aim to compensate for poor learning algorithms by performing a lot of extra computation. Ensemble learning can be used in conjunction with many types of learning algorithms to improve their performance. However, fast algorithms such as decision trees are commonly used, although slower algorithms can benefit from ensemble techniques as well. In this work, we will consider the two most popular categories of ensemble learning, boosting and bagging and as a base learning we will consider trees and discriminant classifiers.

\paragraph{Boosting} Boosting is an iterative process during which the algorithm trains a new classifier in each iteration and adapts it to the final ensemble classifier. The typical idea behind boosting is that each new classifier aims to emphasize the training instances that previous classifiers misclassified. By far, the most common implementation of Boosting is AdaBoost. AdaBoost (short for Adaptive Boosting) combines the outputs of each weak classifier into a weighted sum and provides a stronger classifier by optimizing the weights. The final classifier is proven to converge to a strong one as long as the individual weak classifiers perform at least slightly better than a random guess.

\paragraph{Bagging} Bagging (Bootstrap Aggregating), on the other hand, combines the output of the weak classifiers in the ensemble through voting with equal weight. In order to promote model variance, bagging trains each classifier in the ensemble using a randomly drawn subset of the training set. In the typical case, various subsets of the training set are drawn from the entire training set uniformly and with replacement. Random forests is a special case of Bagging, which uses as a base classifier the decision trees. The main idea of Bagging that trains multiple classifiers on subsets of training data has been also widely used to tackle the class imbalance problem. This problem is common in binary classification, since it is typical to have a large number of negative examples but far fewer positive ones.



\subsubsection{LDA}
\label{sssec:LDA}

Linear Discriminant Analysis (LDA) works in a similar way with SVMs, by attempting to find the separating 
line between the two classes. However, LDA does consider the margin between the classes. More specifically, 
based on the assumption that the covariance matrices of the two classes are equal and have full 
rank ($\boldsymbol{\Sigma_0} = \boldsymbol{\Sigma_1} = \boldsymbol{\Sigma}$), the optimization problem 
degenerates to an analytic form for the optimal $\mathbf{w}$ and $b$ as a function of the 
covariance matrix ($\boldsymbol{\Sigma}$) and the mean ($\boldsymbol{\mu_0}, \boldsymbol{\mu_1}$):
\begin{eqnarray}
\mathbf{w} = \boldsymbol{\Sigma}^{-1} (\boldsymbol{\mu_1} - \boldsymbol{\mu_0}) \\
b = \frac{1}{2}(T-{\boldsymbol{\mu_0}}^T \boldsymbol{\Sigma_0}^{-1} {\boldsymbol{\mu_0}}+
{\boldsymbol{\mu_1}}^T \boldsymbol{\Sigma_1}^{-1} {\boldsymbol{\mu_1}})
\end{eqnarray}
where $T$ is a threshold separating the two classes.

\subsubsection{KNN}
\label{sssec:KNN}

The most popular but also simple algorithm for performing predictions is $k$-nearest neighbors. In the case of KNN,  the mapping function is only approximated locally. This approximation can be a simple voting scheme (i.e. the test instance is assigned to the class most common among its $k$ nearest neighbors) or by assigning weights to the contributions of each neighbor based on their proximity to the test instance.

\subsubsection{Naive Bayes}
\label{sssec:bayesian}

The naive Bayes is a simple probabilistic classifier that attempts to estimate the probability of a class $c_k$ given the features $\mathbf{x}=\{x_1,x_2,...,x_n\}$ (e.g. $p(c_k|\mathbf{x})$ by applying Bayes' theorem. This method is built on a strong (naive) independence assumption, i.e. that the features are independent within each class. Then the probability $p(c_k|\mathbf{x})$ can be calculated as:

\begin{equation}
p(c_k \vert \mathbf{x})=p(c_k \vert x_1, \dots, {x_n}) = \frac{1}{Z} p(c_k) \prod_{i=1}^n p(x_i \vert c_k)
\end{equation}

\noindent where $Z$ is a constant normalization factor.

\section{Data acquisition \& processing}\label{section:dataacquisitionprotocol}

\subsection{Demographics of subjects}

Eleven volunteers participated in this study. They were all present employees of 
Centre for Research and Technology Hellas (CERTH). Specifically, 8 of them were male and 
3 female. Their ages ranged from 25 to 39 years old. All of them were able-bodied subjects 
without any known neuro-muscular or mental disorders. Subjects can also be grouped into 3 categories based on the length and thichkness of their hair (i.e. short hair, regular hair 
and thick hair). Out of the 11 subjects, there are 3 with short hair, 6 with regular hair and the remaining 4 with thick hair. Finally, information about the handedness of each subject was also retained. Table \ref{tbl:subjectDemographics} summarizes the demographics information about the participating subjects, including all the previously discussed information. 

\begin{table}[!ht]
\begin{center}
\caption{General information about the subjects}
\rowcolors{1}{Gray!25}{Gray!5}
\begin{tabular}[c]{cccccc}  \hline
Sub. ID & Age & Gender & Net Size & Hair Type & Handedness \\ \hline \hline
S001 & 24 & Male & Adult Medium & Regular & Right \\ 
S002 & 37 & Male & Adult Small & Regular & Right \\ 
S003 & 39 & Male & Adult Medium & Thick & Right \\ 
S004 & 31 & Male & Adult Medium & Short & Right \\ 
S005 & 27 & Female & Adult Medium & Thick & Left \\ 
S006 & 28 & Female & Adult Medium & Thick & Right \\ 
S007 & 26 & Male & Adult Medium & Regular & Right \\ 
S008 & 31 & Female & Adult Medium & Thick & Right \\ 
S009 & 29 & Male & Adult Medium & Short & Right \\ 
S010 & 37 & Male & Adult Medium & Regular & Right \\ 
S011 & 25 & Male & Adult Medium & Regular & Right \\ \hline
\end{tabular}
\label{tbl:subjectDemographics}
\end{center}
\end{table}

\subsection{Acquisition Setup}

The visual stimuli was projected on a 22'' LCD monitor, with a refresh rate of 60 Hz and 
1680x1080 pixel resolution. The visual stimulation of the experiment was programmed 
in Microsoft Visual Studio 2010 and OpenGL. A graphic card (Nvidia GeForce GTX 860M), 
fast enough to render more frames than the screen can display, was used. Also, the option ``vertical synchronization'' of the graphic card was enabled in order to ensure that only whole frames are seen on screen.

High Dimensional-EEG data were recorded with the EGI 300 Geodesic EEG System (GES 300)~\cite{eeg:300}, using a 256-channel HydroCel Geodesic Sensor Net (HCGSN) and a sampling rate of 250 Hz. The contact impedance at each sensor was ensured to be at most $80K\Omega$ before the initialization of every new session. The synchronization of the stimulus with the recorded EEG signal was performed with the aid of the Stim Tracker model ST - 100 (developed by Cedrus~\cite{stimtracker}), and a light sensor attached to the monitor that added markers (denoted hereafter as Dins) to the captured EEG signal. More specifically, the light sensor was able to detect with high precision the onset of the visual stimuli and place Dins on the EEG signal for as long as the visual stimuli flickered, providing evidence of the lasting period and the frequency of the stimulation. Subsequently, in the offline data processing, these Dins were used to separate the raw signal into the part generated during the visual stimuli and the part  generated during the resting period.

The stimulus of the experiment was one violet box, presented on the center of the monitor, flickering in 5 different frequencies (6.66, 7.50, 8.57, 10.00 and 12.00 Hz). The box 
flickering in a specific frequency was presented for 5 seconds, denoted hereafter as trial,
followed by 5 seconds without visual stimulation before the box appears again flickering 
in another frequency. The background color was black for the whole experiment.

\begin{figure}
\begin{center}
\includegraphics[width=12.0cm]{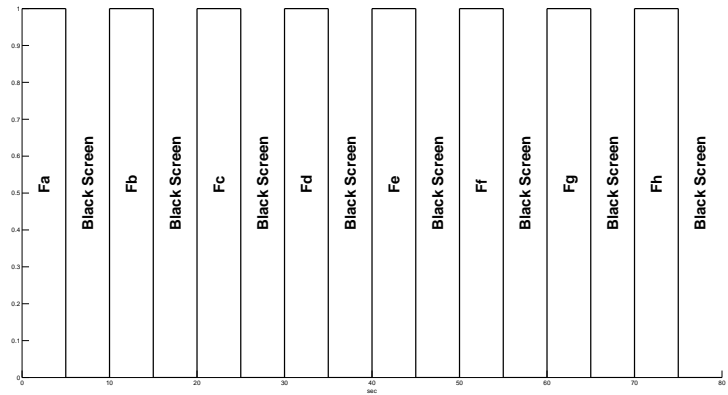}
\caption{Adaptation Experimental Setup: For a period of 80 sec the five stimuli 
are presented randomly to the subject. Between each stimulus a resting period of 5 sec is applied.}
\label{fig:AdaptExpSetup}
\end{center}
\end{figure}

\begin{figure}
\begin{center}
\includegraphics[width=12.0cm]{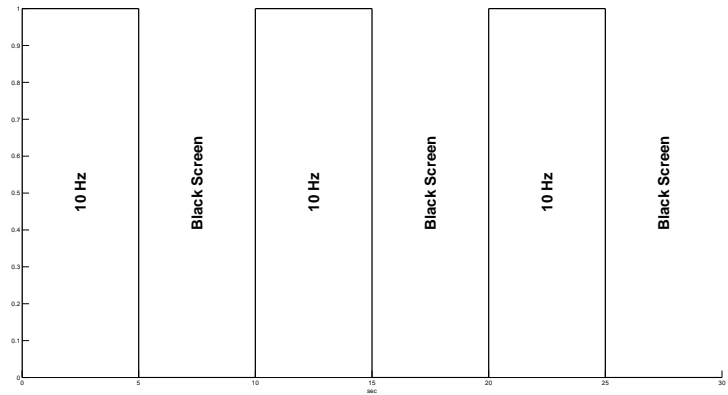}
\caption{Experimental Setup for a particular stimulus. The stimulus is presented for 5sec followed by a resting period of 5sec. The above procedure is applied 3 times. }
\label{fig:StFrExpSetup}
\end{center}
\end{figure}

The experiment process undertaken by each subject was divided into 5 identical sessions. 
Each session was initiated with 100 seconds of resting period where the participant could 
look at the black screen of the monitor without being involved in any activity, followed 
by a 100 seconds of adaptation period (see Fig~\ref{fig:AdaptExpSetup}). 
The adaptation period consisted in the presentation of the 5 selected frequencies in a random way and was considered a crucial part of the process as the subject had the opportunity to familiarize with the visual stimulation. The following 30 
second interval was left for the subject to rest and be prepared for the next trial, which consisted in the presentation of one frequency for 3 times before another 30-second break. 
Every frequency is presented sequential for 3 times and with a resting period of 30 seconds between each trial (see Figure~\ref{fig:StFrExpSetup}). 
Each session eventually includes 23 trials, with 8 of them being part of the adaptation. The entire dataset has been made publicly available in \cite{Georgiadis2016}.    

During the experiment one member of the research stuff was present giving oral instructions 
to the subjects informing them about the resting time they had at their disposal and about 
the time they had (5 seconds) before the resting period would end and the next stimuli would 
appear. In addition, in an effort to minimize the artifacts that could arise by the subject (physiological), the subjects were instructed to limit their movements and try not to swallow or blink during the visual stimulation. Furthermore, the research stuff was responsible for ensuring the correct electrode placement, the movement limitation in the experimental environment and that all mobile phones are switched off. Finally, the participants were cautiously observed and notes were made about unexpected behavior that could lead to existence of artifacts in the acquired signal, in order to use this information later on during the analysis of the classification accuracy.


\subsection{Important notes}\label{subsection:importantnotes}

\noindent\textbf{Eligible signals:} The EEG signal is sensitive to external factors that have to do with the environment or the configuration of the acquisition setup. The research stuff was responsible for the elimination of trials that were considered faulty. As a result, the following sessions were noted and excluded from further analysis: i) S003, during session 4 the stimulation program crashed, ii) S004, during session 2 the stimulation program crashed, and iii) S008, during session 4  the Stim Tracker was detuned. Furthermore, we must also note that subject S001 participated in 3 sessions and subjects S003 and S004 participated in 4 sessions, compared to all other subjects that participated in 5 sessions. As a result, the utilized dataset consists of 1104 trials of 5 seconds each. 

\noindent\textbf{Flickering frequencies:} Usually the refresh rate for an LCD Screen is 60 Hz, creating a restriction to the number of frequencies that can be selected. Specifically, only the frequencies that when divided with the refresh rate of the screen result in an integer quotient could be selected. As a result, the frequencies that could be obtained were the following: 30.00, 20.00, 15.00, 12.00, 10.00, 8.57, 7.50 and 6.66 Hz. In addition, it is also important to avoid using frequencies that are multiples of another frequency, for example making the choice to use 10.00Hz prohibits the use of 20.00 and 30.00 Hz. With the previously described limitations in mind, the selected frequencies for the experiment were: 12.00, 10.00, 8.57, 7.50 and 6.66 Hz.

\noindent\textbf{Stimuli Layout:} In an effort to keep the experimental process as simple as possible, we used only one flickering box instead of more common choices, such as 4 or 5 boxes flickering simultaneously. The fact that the subject could focus on one stimulus without having the distraction of other flickering sources allowed us to minimize the noise of our signals and verify the appropriateness of our acquisition setup. Nevertheless, having concluded to the optimal configuration for analyzing the EEG signals, the experiment will be repeated with more concurrent visual stimulus.

\noindent\textbf{Trial duration:}
The duration of each trial was set to 5 seconds, as this time was considered adequate to allow the occipital part of the brain to mimic the stimulation frequency and still be small enough for making a selection in the context of a brain-computer interface. However, the investigation of the tradeoff between the classification accuracy and the amount of time where the flickering frequency is detected, is included in our immediate plans for future work.

\noindent\textbf{Observed artifacts:} During the stimulus presentation to subject S007 the research stuff noted that the subject had a tendency to eye blink. As a result the interference, in matters of artifacts, on the recorded signal is expected to be high.

\noindent\textbf{Informed consent:} Before the experiment the participants were carefully instructed about the recording procedure and its requirements and were provided with a form of consent to sign after reading it thoroughly. 
After reading the form and listening to our oral instructions, the subjects were motivated to make any questions regarding the procedure in an effort to eliminate misunderstandings about the process. By signing the provided document, the participants stated their voluntary participation in the experiment and their consent to make their data public for research purposes. The entire experimental process has received the approval of the ethics committee of the Centre for Research and Technology Hellas with date 3/7/2015 and for the research grant with number H2020-ICT-2014-644780.  

\subsection{Processing toolbox}\label{subsection:processingtoolbox}

A Matlab toolbox titled ``ssvep-eeg-processing-toolbox'' has been released in GitHub \cite{Liaros2016} along with the dataset in order to setup and perform the experiments described in this paper. It follows a modular architecture that allows the fast execution of experiments of different configurations with minimal adjustments of the code. An experiment pipeline consist of five parts each of them receiving an input from the previous part and providing an output to the next part. The \textbf{Experimenter class} of the toolkit acts as a wrapper on which the parameters of the underlying parts can be specified. The parts that are given as input to an Experimenter object are the following; a) A \textbf{Session} object that is used for loading the dataset and segmenting the signal according to the stimuli markers. It is possible to load a specific session of a subject, all sessions of a specific subject or all the available sessions from all subjects. The output of a Session is a set of trials containing the EEG signal that was recorded during the presentation of the visual stimulus on the screen, annotated with the flashing frequency of the stimulus. b) A \textbf{Preprocessing} object can be used to apply a digital filter on the signals or perform techniques for artifact removal such as AMUSE \cite{Choi:2005}. It is also possible to select a specific channel of the captured signal or to further segment the signal in the time domain. c) A \textbf{Feature Extraction} object receives the processed set of trials as an input and extracts numerical features for representing the signal. Available feature extraction methods include the PWelch power spectral density estimate, Fast Fourier Transform, Discrete Wavelet Transform, etc. d) A \textbf{Feature Selection} object can be also optionally set for receiving the set of feature vectors produced in the feature extraction stage and selecting the ones that are considered the most discriminative based on their entropy or variance. For implementing the process of feature selection, a wrapper class of the FEAST~\cite{Brown:2012} library is included in the toolbox, as well as methods for Principal Component Analysis or Singular Value Decomposition. The input of the feature selection object is the number of retained dimensions and its output is a set of feature vectors with reduced dimensionality. e) A \textbf{Classification} object is the next step of the experiment and includes methods for building a model given a set of annotated feature vectors and for predicting the annotation of unseen feature vectors. Some of the included classifiers are a LIBSVM~\cite{CC01a} wrapper class, AdaBoost and Random Forest. 

The last step before running the experiment is to configure the \textbf{Evaluation} object of the experiment. Currently, there are two methods offered: a) ``Leave one sample out''  trains a classification model with $n-1$ trials and tests the model with the remaining trial. The process is repeated $n$ times until there is an output for each trial; b) ``Leave one subject out'' trains a classification model with the trials belonging to $n-1$ subjects and tests the model with the trials of the remaining subject. The entire process is executed by invoking the ``run'' method of the Experimenter class and when the experiment is finished the results are available via the ``results'' property of the Experimenter class. 

\section{Experimental study}\label{section:experimentalstudy}

As already mentioned, the goal of this paper is to generate a state-of-the-art baseline for SSVEP-based BCIs. Towards this goal, in Section~\ref{section:introduction} we have identified the basic parts of the signal processing module and in Section~\ref{section:algorihmsandmethods} we have listed some of the most prominent algorithms and methods for implementing their functionality. Our goal in this section is to test their performance, so as to optimally tune and configure all different parts of the signal processing module leading to a state-of-the art baseline for SSVEP-based BCIs. Two measures have been used to evaluate the performance of the proposed BCI system, the classification accuracy and the execution time. Finally, all the experiments have been 
performed in an iMac computer with a processor at 3.4GHz (Intel Core i7), memory  at 8GB and an AMD Radeon (1024MB) graphics card.

\subsection{Multiple Parameters Selection} \label{subsection:multiparameterselection}

As made clear in Section~\ref{section:algorihmsandmethods} the efficiency of a SSVEP-based BCI system depends on various different parameters (i.e. filtering, artifact removal, feature extraction, feature selection and classification) that can be implemented by more than one algorithms. This is essentially a multi-objective parameter selection problem that can be tackled with methodologies such as the one presented in \cite{Mussel:2012}, under the restriction that these parameters share some heterogeneity. However, this is not the case for a SSVEP-based BCI system where the regulating parameters are highly heterogeneous. Thus, in our study we have adopted a more empirical approach where each parameter is studied independently by keeping all other parameters fixed. 

More specifically, in order to avoid the tedious process of testing all possible combinations our approach relies on the existence of a default configuration where the value of each parameter has been set arbitrarily (i.e. based on intuition). The role of the default configuration is two-fold; first, to provide the fixed values for all parameters except the one that is being studied and second, to serve as a comparison basis in deciding whether a certain algorithm introduces significant improvements. For example, in deciding which of the feature extraction algorithms presented in Section~\ref{subsection:featureextraction} is the most efficient in the context of SSVEP-based BCIs, we keep all other parameters fixed as dictated by the default configuration and alternate the parameter dealing with feature extraction. If the improvement introduced by the best-performing algorithm is statistically significant with respect to the performance of the default configuration, we retain this algorithm to be part of our state-of-the-art configuration. Although we  acknowledge the fact that the adopted empirical approach may not necessarily lead to the joint optimization of all different parameters, it has been favored over a brute force approach that would make the cost of experiments prohibitive. 

Finally, it is important to note that in studying a certain parameter there are two different optimization process. The first concerns the selection of the best competing algorithm out of the ones presented in Section~\ref{section:algorihmsandmethods}, while the second concerns the optimal tuning of the selected algorithm with respect to its internal variables. Table~\ref{tbl:defaultConfiguration} presents the values that have been selected for the default configuration and unless stated otherwise, are the values that have been used to undertake all experiments described subsequently. 


\begin{table}[!ht]
\begin{center}
\caption{Default Configuration - Parameter Values}
\resizebox{14cm}{!}{
\rowcolors{1}{Gray!25}{Gray!5}
\begin{tabular}[c]{p{10cm}} \hline\hline
\emph{Channel} \\ 
All experiments have been performed by using the raw EEG signal from channel Oz. The place of channel Oz is on the midline
of occipital lobe and, since we study SSVEP responses, it is the logical first choice.\\ \hline\hline
\emph{Duration} \\ 
Duration of used EEG data: all EEG trials (5sec)\\ \hline\hline
\emph{Fitlering} \\ 
The raw EEG signal have been band - pass filtered from 5-48Hz since in this frequency range exists the signal of interest (i.e. the various SSVEP responses). An IIR-Chebyshev I filter is used in the default configuration\\
\hline\hline
\emph{Artifact removal}\\ 
No artifact removal algorithm is used in the default configuration\\ \hline\hline
\emph{Feature extraction}\\ 
For feature extraction the power spectrum of Welch's method is used with the frequency range applied to the entire spectrum and the number of fft set to 512.\\
\hline\hline
\emph{Feature selection} \\ 
In the default configuration we do not apply any feature selection approach/algorithm.\\
\hline\hline
\emph{Classification} \\ 
In the classification step we choose an SVM classifier with a linear kernel and the cost parameter $C$ is set to 1. Since SVMs is essentially a binary classifier the one-vs-all approach is adopted for making this classifier applicable in our multi-class classification problem. \\
\hline\hline
\end{tabular}
\label{tbl:defaultConfiguration}
}
\end{center}
\end{table}

Table~\ref{tbl:performanceDefaultConfiguration} presents the classification accuracy achieved for each subject using the 
default configuration, as well as the mean accuracy for all subjects and the execution time for this configuration. It is important to note that by  execution time we refer to the processing time required by each configuration to reach a decision at the testing phase (i.e. we do not consider the time required for training). In a realistic setting this execution time should be added to the time offered to the subject in order to reach a steady state, which in our case has been set to 5 secs (i.e. the duration of one trial).

By looking at the classification results we may categorize the subjects in three different categories based on their performance: a) \emph{highly-accurate} where we classify all 
subjects with accuracy over 90\% (i.e. SOO1, S009, S010 and S011), b)  \emph{mid-accurate} where we classify all subject 
with accuracy between 60\%-90\% (i.e. S002, S004, S006, S007), and c) \emph{poorly-accurate} where we classify all subjects 
with accuracy below 60\% (i.e. S003, S005, S008). By considering these categories in conjunction with the information of
Table~\ref{tbl:subjectDemographics} we may reach the following conclusions. All subjects in the \emph{highly-accurate} 
and \emph{mid-accurate} categories have either short or regular hair (with the only exception of S006 that has thick hair), 
while all subjects in the \emph{poorly-accurate} category appears to have thick hair. This observation verifies the knowledge 
obtained from literature that thick hair constitute a series obstacle in acquiring noise-free EEG signals. Another interesting 
remark concerns the mid-to-poor accuracy of subject S007 that has been observed to excessively blink during the execution of 
the experiment (see Section~\ref{subsection:importantnotes}). Finally, the last remark concerns subject S005, which is the 
only left-handed subject participating in our study. 

\begin{table}[!ht]
\begin{center}
\caption{Classification accuracy using the default configuration.}
\resizebox{5cm}{!}{
\rowcolors{1}{Gray!25}{Gray!5}
\begin{tabular}[c]{cc}  \hline
\emph{Subject ID} & \emph{Accuracy} \\ \hline\hline
S001    & 98.55     \\
S002     & 87.82     \\
S003   & 34.78     \\
S004    & 77.17     \\
S005  & 30.43     \\
S006  & 86.08      \\
S007    & 60.00          \\
S008   & 31.88     \\
S009   & 100.0         \\
S010  & 92.17     \\
S011     & 98.26     \\ \hline\hline
\textbf{Mean Accuracy}     & 72.47 \\
\textbf{Time (msec)} & 5
\end{tabular}
\label{tbl:performanceDefaultConfiguration}
}
\end{center}
\end{table}

\subsection{Evaluation protocol}\label{subsection:evaluationprotocol}
In order to perform the necessary comparisons across the different configurations, we need to determine an evaluation protocol that will allow us to obtain a performance indicator for each test. This procedure falls into the model selection problem, since we have several candidate models and we wish to choose the best of them with respect to an optimality criterion. A popular framework to choose a model among a set of candidate models is the Cross - Validation (CV) approach \cite{Arlot:2010}. The general idea of CV is to split the available dataset in two parts, the training set and the testing set. Then, a learning procedure for each candidate model is performed using the training set, in order to learn/estimate the free parameter of the  model. After that, the performance of each candidate model is evaluated with respect to the testing set. Finally, the model with the best performance in the testing set is chosen. The different variations of CV framework are related with the splitting procedure and the metric used to quantify the performance of the model.

In our study, we choose to employ a CV approach where the 
splitting procedure is performed on the basis of subjects \cite{Rice:1991,Xu:2012,Arlot:2010} and the performance is calculated
by the accuracy of the classifier. This approach is called Leave-One-Subject-Out (LOSO) and has been previously used to discriminate between sensorimotor rhythms (two-class problem) in a BCI system \cite{Fazli:2009}.
As the name indicates the LOSO-CV approach suggests to leave the data from one subject out of the training phase and use them only in the testing phase of the experiment. This splitting is very important for BCI experiments since it provides us the ability to construct general purpose systems, free of the necessity to perform subject-specific training prior to operation. 

The LOSO - CV approach treats the EEG data quite different than a classification scheme that 
relies on subject-specific classifiers and work independently. More specifically, in the case 
of a subject-specific classifier where both the training and testing sets are generated from the 
same subject, we are unable to account for the between - subject variability. To avoid subject-dependency 
one could train/test a classifier in the EEG data from all subjects belonging to the group using for 
example a 10-fold CV approach. But in this approach the data from one subject are included in both the 
training and testing set, which reduces the withing subject variability of the resulting classifier. 
To preserve the within subject variability and at the same time to take into account the between subjects 
variability we use a CV approach where the splitting procedure is performed with respect to the subjects 
and not to the samples, such as in the case of LOSO-CV.


\subsection{Signal Filtering}
In this series of experiments our goal is to evaluate the two basic families of filters, FIR and IIR, with 
respect to the classification accuracy, and to highlight the usefulness of filtering process. Using the 
default configuration of our methodology we perform a comparison between filtered and raw data, as well as 
between various filters.
Five basic types of filters are used as reported 
in Table \ref{tbl:filteringRes}. These filters are designed using the Matlab Signal Processing Toolbox 
(tool \texttt{filterbuilder}\footnote{http://www.mathworks.com/help/dsp/ref/filterbuilder.html}). 
In all cases the EEG trials have been bandpass filtered from 5-48Hz. Also, before the filtering, the EEG trials
have been normalized to zero mean. The filters specifications 
are described in Table \ref{tbl:filterSpec}. In addition to the above specifications the order of 
FIR filters have been restricted to 400,
while the order of IIR filters has been defined by the tool \texttt{filterbuilder}.
In Table \ref{tbl:filteringRes} we depict the obtained results for this series of experiments.
At first, we can see that filtering the raw signal is absolutely necessary since 
we achieve a classification accuracy 30\% larger than without the filtering procedure.
Furthermore, between the various filters we can see that the IIR filters is slightly better (0.5\%) than FIR, 
with the IIR-Elliptic filter achieving the best performance. This difference can be explained probably by the way that 
each filter treats the ripples in the passband and stopband. With respect to the execution time we see that all configurations (i.e. different filter in this case) provides with similar results. Based on the obtained results the IIR-Elliptic filter has been chosen for our state-of-the-art configuration. 

\begin{table}[!ht]
\begin{center}
\caption{Specifications of filters}
\resizebox{11cm}{!}{
\rowcolors{1}{Gray!25}{Gray!5}
\begin{tabular}[c]{cccc}  \hline
Frequency Specifications (Hz)& & & \\
Stopband Frequency 1 & 4 & Passband Frequency 1 & 5 \\
Passband Frequency 2 & 48 & Stopband Frequency 2 & 50 \\
Magnitude Specifications (dB)& & & \\
Stopband Attenuation 1 & 60 & Passband Ripple & 1 \\
Stopband Attenuation 2 & 60 & & \\
\end{tabular}
\label{tbl:filterSpec}
}
\end{center}
\end{table}

\begin{table}[!ht]
\begin{center}
\caption{Classification accuracy using various filters.}
\resizebox{\textwidth}{!}{
\rowcolors{1}{Gray!25}{Gray!5}
\begin{tabular}[c]{p{1.5cm}p{1.5cm}p{1.5cm}p{1.5cm}p{1.5cm}p{1.5cm}p{1.5cm}}  \hline
\emph{Sub. ID} & FIR Equirrible	& FIR Least Squares	
			 	& IIR  Chebyshev I & 
            IIR  Chebyshev II	& IIR Elliptic &Raw EEG trials\\ \hline\hline
S001          & \textbf{100.0}         & 98.55     & 98.55     & \textbf{100.0}         & \textbf{100.0}         & 75.36     \\
S002          & 86.95     & 87.82     & 87.82     & \textbf{90.43}     & 89.56     & 33.91      \\
S003          & 33.33     & 30.43     & \textbf{34.78}     & 33.33     & 33.33     & 24.63     \\
S004          & 79.34     & \textbf{80.43}     & 77.17     & 79.34     & 78.26     & 39.13     \\
S005          & 21.73     & 25.21     & \textbf{30.43}     & 28.69     & 29.56     & 20.86     \\
S006          & \textbf{86.08}      & 84.34     & \textbf{86.08}      & \textbf{86.08}      & \textbf{86.08}      & 39.13     \\
S007          & 64.34     & \textbf{65.21}     & 60.00          & 60.86     & \textbf{65.21}     & 34.78     \\
S008          & 33.33     & \textbf{36.23}     & 31.88     & 33.33     & 34.78     & 21.73     \\
S009          & \textbf{100.0}         & \textbf{100.0}         & \textbf{100.0}         & \textbf{100.0}         & \textbf{100.0}         & 75.65     \\
S010          & \textbf{93.04}     & 91.30     & 92.17     & 92.17     & 91.30     & 40.00          \\
S011          & \textbf{98.26}     & \textbf{98.26}     & \textbf{98.26}     & \textbf{98.26}     & \textbf{98.26}     & 87.82     \\ \hline
\textbf{Mean Acc.}  & 72.40 & 72.52 & 72.47 & 72.95 & \textbf{73.30} & 44.82 \\
\textbf{Time (msec)} & 6	& 5 &	5 &	5	& 5	& 3
\end{tabular}
\label{tbl:filteringRes}
}
\end{center}
\end{table}

\subsection{Artifact Removal}
Our objective in this section is to compare the artifact removal algorithms presented in Section~\ref{subsubsec:artifactremoval}, as 
well as to verify the positive impact of these algorithms in the final classification result. In this direction, 
we have performed experiments using AMUSE and FastICA. More specifically, both algorithms have been applied on a 
trial basis (i.e. using the 5 secs of EEG signal captured in each trial) and using all 256 channels, so as to 
generate the necessary components. Subsequently, the visual inspection of the data resulted in the identification 
of the components that could potentially include artifacts. Finally, different combinations of these components 
were removed before reconstructing the EEG signal using the remaining components. The components that have been 
retained for the signal reconstruction and the resulting accuracy of each method can be seen in Table~\ref{tbl:artifactRemoval}.

\begin{table}[!ht]
\begin{center}
\caption{Results using Artifact Removal Methods}
\resizebox{\textwidth}{!}{
\rowcolors{1}{Gray!25}{Gray!5}
\begin{tabular}[c]{p{0.75cm}p{0.85cm}p{0.85cm}p{0.85cm}p{0.85cm}p{0.85cm}p{0.85cm}p{0.85cm}p{0.85cm}p{0.85cm}p{0.9cm}p{0.9cm}p{0.9cm}p{0.9cm}}  \hline
\emph{Sub. ID} & \multicolumn{9}{c}{\cellcolor{gray!25} \emph{AMUSE} } &\multicolumn{4}{c}{\cellcolor{gray!25} \emph{ICA}} \\ \hline
comp & 2-256 & 2-255       & 2-252 & 2-247 & 10-256 & 15-256 & 20-256 & 15-255 & 15-252 & 120-256 & 130-256 & 140-256 & 150-256 \\ \hline\hline
S001    & 98.55 & 98.55       & 97.10 & 97.10 & \textbf{100.0}  & \textbf{100.0}  & 91.30  & \textbf{100.0}  & \textbf{100.0} \vline & 17.39   & 30.43   & 21.73   & 21.73   \\
S002    & 86.95 & 82.60       & 84.34 & 83.47 & 93.91  & 93.04  & 84.34  & 92.17  & \textbf{95.65} \vline& 21.73   & 23.47   & 22.60   & 19.13   \\
S003    & 33.33 & 42.02       & 40.57 & 42.02 & 33.33  & 46.37  & 39.13  & \textbf{47.82}  & 40.57 \vline & 15.94   & 21.73   & 21.73   & 23.18   \\
S004    & 75.00 & 76.08       & 75.00 & 75.00 & 71.73  & 76.08  & 69.56  & \textbf{77.17}  & 76.08  \vline & 16.30   & 21.73   & 21.73   & 23.91   \\
S005    & 26.95 & 24.34       & 24.34 & 20.86 & 23.47  & 25.21  & 20.86  & 24.34  & \textbf{32.17} \vline & 21.73   & 16.52   & 15.65   & 18.26   \\
S006    & \textbf{83.47} & 80.00       & 69.56 & 66.95 & 75.65  & 70.43  & 56.52  & 67.82  & 67.82 \vline & 20.86   & 21.73   & 21.73   & 20.86   \\
S007    & 69.56 & 74.78       & 84.34 & 88.69 & 92.17  & 90.43  & 72.17  & 90.43  & \textbf{95.65} \vline & 19.13   & 21.73   & 21.73   & 21.73   \\
S008    & 31.88 & 23.18       & \textbf{34.78} & 24.63 & 33.33  & 28.98  & \textbf{34.78}  & 30.43  & 28.98 \vline & 14.49   & 20.28   & 20.28   & 27.53   \\
S009    & 99.13 & 99.13       & 99.13 & 99.13 & \textbf{100.0}  & \textbf{100.0}  & 98.26  & \textbf{100.0}  & \textbf{100.0} \vline & 23.47   & 21.73   & 33.04   & 32.17   \\
S010    & \textbf{98.26} & \textbf{98.26}       & \textbf{98.26} & 95.65 & 96.52  & 94.78  & 93.91  & 93.04  & 95.65 \vline & 14.78   & 20.86   & 20.86   & 20.00   \\
S011    & \textbf{98.26} & \textbf{98.26}     & \textbf{98.26} & \textbf{98.26} & \textbf{98.26}  & 93.04  & 85.21  & 93.91  & 93.91  \vline & 14.78   & 31.30   & 20.86   & 22.60   \\\hline\hline 
\textbf{Mean Acc.}  & 72.85 & 72.47 & 73.24 & 71.98 & 74.40  & 74.40  & 67.82  & 74.28  & \textbf{75.13}  \vline & 18.24   & 22.87   & 22.00   & 22.83  \\
\textbf{Time (msec)} & 81	&73&	70&	68&	70&	69&	69&	70&	68 &5660 & 5660& 5660&5660
\end{tabular}
\label{tbl:artifactRemoval}
}
\end{center}
\end{table}

It is evident from the results of Table~\ref{tbl:artifactRemoval} that AMUSE is a reliable artifact 
removal method as the mean accuracy is improved in most of the cases. The highest accuracy is achieved 
(75.13\%) when the last five and the first fifteen components produced by AMUSE are excluded, with an improvement 
of approximately 2.5\% compared to the default configuration. Furthermore, it is important to notice that 
the accuracy for Subject S007, a Subject that was observe to blink excessively during the recordings 
(see Section~\ref{subsection:importantnotes}), was improved up to 30\% verifying the effectiveness of AMUSE 
in removing the artifacts generated from eye-blinks. On the contrary, the use of FastICA didn't manage to 
introduce any improvements compared to the default configuration, regardless of the utilized components. 
Also, we can see that the execution time of AMUSE is much smaller than that of FastICA. Nevertheless, the 68 msecs required by AMUSE to remove the artifacts from the signal is a significant increase in the total execution time, compared to the 5 msecs of the default configuration. Based on these results, the use of AMUSE combined with the removal of the first 15 and the last 4 components, 
was incorporated as part of our optimal configuration. 

\subsection{Feature Extraction}


In these series of experiments our goal is to compare the discrimination ability of 
the feature extraction methods presented in Section~\ref{subsection:featureextraction} on our dataset. 
For all steps of the procedure, except the Feature Extraction step, the default configuration has been used. 
The tuning parameters for each of the feature extraction methods have been obtained through a set of preliminary 
experiments and based on trial-and-error. Table~\ref{tbl:featureExtractionResults} depicts the classification 
accuracy achieved using each of the feature extraction methods along with its tuning parameters.  

We can see the superiority of Welch method as a feature extractor with respect to SSVEP experiments. This can 
be attributed to the averaging effect of Welch methods that seem to favor the robustness of the classification. 
Also, we can see that all Fourier based methods achieve better results than the Discrete Wavelet Transform. 
This does not exclude the presence of non-stationarities in the EEG signal, but it is an indication that the
brain’s synchronization with the stimulus produces a stationary process. This 
can be also justified from the fact that the duration of EEG trials was 5sec, a 
time window large enough for the brain to come into a steady state situation 
with respect to the visual stimulus. 
However, not in all cases the Welch method provides the best classification results. For example,
in the case of subject S005 the wavelet transform provides us with better classification accuracy. It is worth to notice here that the Goertzel algorithm needs more processing time than the others
approaches.
Based on the results of Table~\ref{tbl:featureExtractionResults}, the PWelch method qualifies as the best feature 
extraction method to be included in our optimal configuration. 


\begin{table}[!ht]
\begin{center}
\caption{Classification Accuracy using various Feature Extraction Methods.}
\resizebox{15cm}{!}{
\rowcolors{1}{Gray!25}{Gray!5}
\begin{tabular}[c]{p{1.3cm}p{1.5cm}p{1.5cm}p{1.5cm}p{1.5cm}p{1.5cm}p{1.6cm}}  \hline
\emph{Sub. ID} & \emph{PWelch} & \emph{Periodogram} & \emph{PYAR} & \emph{DWT} & \emph{STFT} & \emph{Goertzel} \\
 & Nfft:512 Fr. range: 0-125Hz  & Nfft:512  Fr. range: 0-125Hz  & Nfft:512  Fr. range: 0-125Hz  AR:20 & Wav.:db1  Dec.:5 & 
 Nfft:512  Fr. range: 0-125Hz  &  Fr. range: 0-125Hz 
\\ \hline\hline
S001          & \textbf{98.55}     & 81.15     & 84.05      & 69.56     & 94.20     & 86.95     \\
S002          & \textbf{87.82}     & 60.86     & 50.43     & 37.39     & 70.43     & 57.39     \\
S003          & 34.78     & 27.53     & 20.28     & 33.33     & 30.43     & \textbf{36.23}     \\
S004          & \textbf{77.17}     & 71.73     & 45.65     & 34.78     & 54.34     & 72.82     \\
S005          & 30.43     & 26.08      & 26.95     & \textbf{34.78}     & 33.04     & 23.47     \\
S006          & \textbf{86.08}      & 66.95     & 39.13     & 60.86     & 61.73     & 66.08      \\
S007          & \textbf{60.00}          & 50.43     & 38.26     & 35.65     & 50.43     & 45.21     \\
S008          & \textbf{31.88}     & 24.63     & 30.43     & 23.18     & 23.18     & 23.18     \\
S009          & \textbf{100.0}         & 90.43     & 73.04     & 81.73     & 97.39     & 86.08      \\
S010          & \textbf{92.17}     & 60.86     & 55.65     & 31.30     & 76.52     & 68.69     \\
S011          & \textbf{98.26}     & 97.39     & 72.17     & 73.91      & 98.26     & 95.65     \\\hline\hline 
\textbf{Mean Acc.} & \textbf{72.47} & 59.82 & 48.73 & 46.95 & 62.72 & 60.16 \\
\textbf{Time (msec)} & 5	& 2	& 2 &	2 &	3	& 21 \\
\end{tabular}
\label{tbl:featureExtractionResults}
}
\end{center}
\end{table}

As depicted in Table~\ref{tbl:featureExtractionResults} the tuning parameters of PWelch has been set to Nfft:512 
and Fr. range: 0-125Hz. Given that PWelch has been selected as the best-performing candidate, the next series of 
experiments aimed at further optimizing the tuning parameters of PWelch so as to increase the classification 
accuracy. More specifically, the number of FFT points, the length of the segments and the overlap between segments 
have been checked using a grid-search approach. The frequency range has been kept fixed at 0-125Hz. The overall tuning procedure is described in Algorithm \ref{alg1} and the accuracy of the 10 best configurations resulting from this procedure is depicted in Table \ref{tbl:tuningWelch}. 
Also, in this table we provide the accuracy corresponding to the default values of the above parameters that have 
been used in Table~\ref{tbl:featureExtractionResults}. As we can see
by further tuning the PWelch method we manage to obtain an increase in accuracy around 1\%. However, this increase on classification is coming at the expense of 1 additional msec in the total execution time. From the above results it is evident that an increase in the overlap between segments provides us with better accuracy (although it requires more processing time), probably due to the fact that the estimated frequencies are more accurate.  



\begin{algorithm}
\caption{Tuning PWelch method}
\label{alg1}
\begin{algorithmic}
\STATE {$\text{nfft} = [128, 256, 512, 1024, 2048]$} \COMMENT{number of FFT points}
\STATE {$\text{win\_len} = [125, 200, 250, 300, 350, 400, 450, 500]$} \COMMENT{length of each segment}
\STATE {$\text{over\_len} = [0.25, 0.5, 0.75, 0.9]$} \COMMENT{percentage of overlap}
\FOR {$i=1$ to $length(\text{nfft})$}
 \FOR{ $j=1$ to $length(\text{win\_len})$}
  \FOR {$k=1$  to $length(\text{over\_len})$}
  	\STATE {Apply the proposed procedure to obtain the classification accuracy}
  \ENDFOR
 \ENDFOR
\ENDFOR
\end{algorithmic}
\end{algorithm}

\begin{table}[!ht]
\begin{center}
\caption{Tuning PWelch method}
\rowcolors{1}{Gray!25}{Gray!5}
\begin{tabular}[c]{ccccc} \hline
Number of FFT points & Segment length & Overlap & Mean Acc. & Time (msec)\\ \hline \hline
512  & 350 & 0.75 & \textbf{73.32} & 6\\
512  & 400 & 0.75 & 73.21 & 5\\
2048 & 200 & 0.9  & 73.19 & 23 \\
2048 & 250 & 0.75 & 73.17 & 9\\
1024 & 350 & 0.75 & 73.08 & 6\\
512  & 250 & 0.9  & 73.05 & 17\\
1024 & 250 & 0.9  & 73.04 & 18\\
2048 & 250 & 0.9  & 73.01 & 18	\\
512  & 300 & 0.9  & 72.92 & 14\\
512  & 250 & 0.75 & 72.90 & 9\\\hline\hline
\multicolumn{5}{c}{Default Values for PWelch used in Table~\ref{tbl:featureExtractionResults}} \\ \hline \hline
512 & 156 & 0.5 & \textbf{72.47}& 5 \\
\end{tabular}
\label{tbl:tuningWelch}
\end{center}
\end{table}


\subsection{Feature Selection}

In this section we experimentally examine the effectiveness of the feature selection methods 
presented in Section~\ref{subsection:featureselection} to increase the discrimination capacity 
of the feature space and lead to better classification results. Aiming in the investigation of 
the features selection parameter, the default configuration was used to set the values for all 
other parameters of signal processing, alternating only between the different option for feature 
selection. Table~\ref{tbl:featureSelectionResults} presents the comparative results among the entropy-based 
scoring criteria presented in Section~\ref{subsubsection:entropyFeatureSelection} and the data projection 
methods presented in Section~\ref{subsubsection:dataprojectionFeatureSelection}. In order to ensure a fair 
comparison between the different criteria, we choose to select 80 features out of the original feature set (257 features) in 
all cases, except from the case of Jcmi where the maximum number of features returned by this method is 24. 
It is evident from Table~\ref{tbl:featureSelectionResults} that by performing feature selection using SVD 
achieves the highest classification accuracy among all different candidates. We 
can also observe that most of the employed feature selection approaches fail to introduce significant improvements 
compared to the default configuration (where no feature selection is applied), or even decrease the 
mean accuracy over all subjects. Finally, with respect to the execution 
time we can see that all configurations (i.e. different feature selection methods) provides with similar results.

\begin{table}[!ht]
\begin{center}
\caption{Classification Accuracy using various Feature Selection Methods.}
\resizebox{\textwidth}{!}{
\rowcolors{1}{Gray!25}{Gray!5}
\begin{tabular}[c]{p{0.75cm}p{0.7cm}p{0.7cm}p{0.7cm}p{0.7cm}p{0.7cm}p{0.7cm}p{0.7cm}p{0.7cm}p{0.7cm}p{0.6cm}p{0.7cm}p{0.7cm}p{0.7cm}}  \hline
\emph{Sub. ID} & \emph{Jmim} & \emph{Jmifs} & \emph{Jjmi} & \emph{Jcmi} & \emph{Jmrmr} & \emph{Jcmim} & \emph{Jicap} & \emph{Jcife} & \emph{Jdisr} & \emph{Jbg} & \emph{Jcond} & \emph{PCA} & \emph{SVD} \\
 & d:80 & d:80 & d:80 & d:24 & d:80 & d:80 & d:80 & d:80 & d:80 & d:80 & d:80 & d:80 & d:80\\ \hline\hline
S001    & 98.55 & 98.55 & 98.55 & \textbf{100.0}   & 98.55 & 98.55 & 98.55 & 98.55 & 98.55 & 98.55 & 98.55 & 98.55 & 97.10 \\
S002    & 86.95 & 86.95 & 88.69 & 66.95 & 89.56 & 86.95 & 88.69 & 88.69 & 86.95 & 86.95 & 86.95 & 87.82 & \textbf{92.17} \\
S003    & 34.78 & 34.78 & 36.23 & 27.53 & 34.78 & 31.88 & 31.88 & 33.33 & 34.78 & 34.78 & 37.68 & 34.78 & \textbf{37.68} \\
S004    & 77.17 & 77.17 & 77.17 & 67.39 & 72.82 & 76.08 & 76.08 & 77.17 & 75.00    & 77.17 & 76.08 & 77.17 & \textbf{78.26} \\
S005    & \textbf{33.04} & \textbf{33.04} & 30.43 & 25.21 & 26.95 & \textbf{33.04} & 32.17 & 31.30 & 31.30 & \textbf{33.04} & \textbf{33.04} & 30.43 & 24.34 \\
S006    & 86.08 & 86.08 & 85.21 & 71.30 & 84.34 & 85.21 & 85.21 & \textbf{86.95} & 85.21 & 86.08 & 85.21 & 86.08 & 86.08 \\
S007    & 59.13 & 59.13 & 59.13 & 66.95 & 60.00    & 60.00    & 58.26 & 60.86 & 60.00    & 59.13 & 61.73 & 60.00    & \textbf{65.21} \\
S008    & 33.33 & 33.33 & 33.33 & 23.18 & 40.57 & 31.88 & 33.33 & 33.33 & 33.33 & 33.33 & \textbf{34.78} & 31.88 & 28.98 \\
S009    & \textbf{100.0}   & \textbf{100.0}   & \textbf{100.0}   & 97.39 & \textbf{100.0}   & \textbf{100.0}   & \textbf{100.0}   & \textbf{100.0}   & \textbf{100.0}   & \textbf{100.0}   & \textbf{100.0}   & \textbf{100.0}   & \textbf{100.0}   \\
S010    & 92.17 & 92.17 & 91.30 & 60.00    & 89.56 & 91.30 & 91.30 & 89.56 & 89.56 & 92.17 & 89.56 & 92.17 & \textbf{95.65} \\
S011    & \textbf{98.26} & \textbf{98.26} & \textbf{98.26} & 94.78 & \textbf{98.26} & \textbf{98.26} & \textbf{98.26} & \textbf{98.26} & \textbf{98.26} & \textbf{98.26} & 97.39 & \textbf{98.26} & \textbf{98.26} \\ \hline\hline 
\textbf{Mean Acc.} & 72.68 & 72.68 & 72.57 & 63.70 & 72.31 & 72.10 & 72.16 & 72.54 & 72.08 & 72.68 & 72.81 & 72.47 & \textbf{73.06} \\
\textbf{Time (msec)} & 5	&6	&5	&5&	5	&5	&6	&6	&6&	6	&6	& 5	& 5
\end{tabular}
\label{tbl:featureSelectionResults}
}
\end{center}
\end{table}

Having concluded SVD to be the best method for feature selection, we wanted to further examine the number 
of selected features and its impact on the overall classification accuracy. In this direction, a wide range 
of values for the number of features were tested that started using one quarter of the features and ending 
up using half of them. As we can see in Table~\ref{tbl:tuningICAP}, by setting the number of selected features 
to 90 produces the best accuracy results for SVD. Furthermore, increasing the number of selected features 
does not provide us with better accuracy. 
Compared to the default configuration we can see that the improvement derives from slight modifications in the accuracy 
for nearly half of the subjects and doesn't allow for any subject-specific interpretation. 
Finally, based on the obtained results the feature selection method relying on SVD and the dimensionality of 90 (out of 257) features were selected to become part of our optimal configuration. 

\begin{table}[!ht]
\begin{center}
\caption{Tuning the number of selected feature for SVD}
\resizebox{16cm}{!}{
\rowcolors{1}{Gray!25}{Gray!5}
\begin{tabular}[c]{ccccccccc}
No. of Comp. & 65 & 70 & 75 & 80 & 85 & 90 & 95 & 100 \\ \hline\hline
S001    & \textbf{98.55}     & \textbf{98.55}     & 97.10     & 97.10     & \textbf{98.55}     & \textbf{98.55}     & \textbf{98.55}     & \textbf{98.55}     \\
S002    & 86.08     & 87.82     & 86.08     & \textbf{92.17}     & 86.08     & 86.95     & 87.82     & 87.82     \\
S003    & 39.13     & 37.68     & 36.23     & \textbf{37.68}     & 34.78     & 34.78     & 34.78     & 34.78     \\
S004    & 76.08     & 75.00     & 72.82     & \textbf{78.26}     & 73.91     & 77.17     & 78.26     & 77.17     \\
S005    & 25.21     & 28.69     & 28.69     & 24.34     & 30.43     & 29.56     & \textbf{32.17}     & 30.43     \\
S006    & 80.86     & 82.60     & 80.86     & 86.08     & 86.95     & \textbf{87.82}     & 86.95     & 86.08     \\
S007    & 59.13     & 62.60     & 66.08     & \textbf{65.21}     & 60.86     & 62.60     & 58.26     & 60.00     \\
S008    & 28.98     & 28.98     & \textbf{31.88}     & 28.98     & \textbf{31.88}     & 30.43     & 30.43     & \textbf{31.88}     \\
S009    & 99.13     & 99.13     & \textbf{100.0}     & \textbf{100.0}     & \textbf{100.0}     & \textbf{100.0}     & \textbf{100.0}     & \textbf{100.0}     \\
S010    & 87.82     & 91.30     & 93.91     & 95.65     & 97.39     & \textbf{98.26}     & 96.52     & 92.17     \\
S011    & \textbf{98.26}     & 97.39     & 97.39     & \textbf{98.26}     & \textbf{98.26}     & \textbf{98.26}     & \textbf{98.26}     & \textbf{98.26}     \\\hline\hline
\textbf{Mean Acc.} & 70.84 & 71.79 & 71.91 & 73.06 & 72.64 & \textbf{73.12} & 72.91 & 72.47 \\
\textbf{Time (msec)} & 5 &	5 &	5 &	5	& 5	& 5 & 5 & 5 
\end{tabular}
\label{tbl:tuningICAP}
}
\end{center}
\end{table}

\subsection{Classification}

The objective of this section is to perform a series of experiments in order to decide which 
classifier is more suitable for SSVEP data analysis with respect to both accuracy and time. 
At first, we test various popular machine learning algorithms; more specifically, we opt to experiment
with SVMs, ensemble learning (using either Adaboost or bagging for the ensemble and either trees or 
discriminant classifier as the basis for the ensemble), linear discriminant analysis (LDA), KNN and 
Naive Bayes. For SVMs, the libSVM library was used~\cite{CC01a} with the library's default parameters 
(i.e. linear kernel and C=1). For the rest of the algorithms, we have relied on the implementations 
of MATLAB's Statistics and Machine Learning toolbox using the default parameters for each classification 
scheme, which can be found at MATLAB's online manual\footnote{http://www.mathworks.com/help/stats/index.html}. 
Ensemble learning was examined using Adaboost or Bagging as the ensemble creation method and discriminant 
or tree-based classifiers as the base learning algorithm. All the ensembles were created using 100 weak 
classifiers. The results can be seen in Table~\ref{tbl:comparingClassifiers}. It is evident that SVMs 
provide the most robust performance at reasonable computational cost. It is also worth noting that SVMs 
outperform all other classifiers for 10 out of the 11 subjects in terms of classification accuracy, while at the same time being one of most computationally efficient algorithms. 



In the process of tuning the SVMs, it is important to select the appropriate kernel, which will transform 
the input data into a space where they are separable. In order to find the appropriate kernel, we explore 
a wide variety of them ranging from popular kernels (e.g RBF), to more target specific ones (e.g. correlation). 
First, we test the popular kernels, i.e. linear, RBF and chi-square. Second, we take the formula of the RBF 
kernel (i.e. $K(x,y)=\exp^{-\gamma \cdot d(x,y)^2}, \gamma > 0$, where $d(x,y)$ is the euclidean distance for 
the RBF kernel, and replace it with other popular distance metrics (standardized euclidean, 
cityblock, minkowski and Chebyshev). Finally, we also evaluate similarity metrics that could be more fitting 
to the EEG domain, i.e. a signal based domain (cosine similarity, cross correlation and Spearman correlation). 
The results can be seen in Table~\ref{tb:comparingKernels}. We can see that the Spearman kernel outperforms 
significantly all the other kernels in 6 out of 11 cases as well as in average numbers, 
providing the best results. 
By taking a closer 
look we can see that the average score is primarily boosted by the improvement introduced for subjects 
S003 and S008. 
The superiority of this kernel can be explained if we consider that the Spearman's rank correlation is a 
measure of statistical dependence. Unlike the distance based kernels, which are based on the difference 
between the values of the features,  Spearman's rank correlation finds monotonic relations between the 
features by treating them as ''sequences''. In this way it ignores the absolute differences between the 
values and relies only on the statistical dependence of the features. This can be particularly useful in 
the case of different subjects, where each one may have a different reaction to the presence (or absence) 
of external stimuli in terms of absolute values but still produce correlated spectra of signals. Finally, 
based on the aforementioned results, the SVM classifier using a kernel based on Spearman correlation was 
selected to become part of our optimal configuration.

\begin{table}[!ht]
\centering
\caption{Results using various classifiers}
\rowcolors{1}{Gray!25}{Gray!5}
\label{tbl:comparingClassifiers}
\resizebox{\textwidth}{!}{
\begin{tabular}
{p{1cm}p{0.8cm}p{1.1cm}p{1.1cm}p{1.1cm}p{1.1cm}p{1.1cm}p{0.9cm}p{0.9cm}p{0.9cm}} 
\emph{Sub. ID} & \emph{SVMs} & \emph{Decision Trees} & \multicolumn{4}{c}{\cellcolor{gray!25} \emph{Ensemble - \#Classifiers = 100}} & \emph{LDA} & \emph{KNN} & \emph{Naive Bayes} \\
  & & & Boost, Discr. & Boost, Trees & Bag, Discr. & Bag, Trees &  &  &\\
  \hline\hline
S001    & \textbf{98.55} & 69.56 & 97.10 & 50.72 & 94.20 & 91.30 & 95.65 & 92.75 & 68.11 \\
S002    & \textbf{87.82} & 65.21 & 68.69 & 29.56 & 73.04 & 82.60 & 73.04 & 38.26 & 31.30 \\
S003    & \textbf{34.78} & 28.98 & 28.98 & 24.63 & 30.43 & 34.78 & 27.53 & 28.98 & 20.28 \\
S004    & \textbf{77.17} & 50.00 & 66.30 & 47.82 & 69.56 & 71.73 & 64.13 & 53.26 & 31.52 \\
S005    & \textbf{30.43} & 24.34 & 20.86 & 24.34 & 20.86 & 25.21 & 19.13 & 20.00 & 21.73 \\
S006    & \textbf{86.08} & 50.43 & 74.78 & 46.08 & 80.00 & 82.60 & 79.13 & 34.78 & 22.60 \\
S007    & \textbf{60.00} & 58.26 & 43.47 & 56.52 & 53.04 & 66.95 & 48.69 & 59.13 & 31.30 \\
S008    & \textbf{31.88} & 20.28 & 21.73 & 21.73 & 20.28 & 20.28 & 23.18 & 31.88 & 31.88 \\
S009    & \textbf{100.0} & 77.39 & 99.13 & 72.17 & 97.39 & 98.26 & 98.26 & 71.30 & 40.86 \\
S010    & \textbf{92.17} & 52.17 & 79.13 & 30.43 & 77.39 & 86.95 & 78.26 & 33.91 & 24.34 \\
S011    & 98.26 & 63.47 & \textbf{99.13} & 46.08 & 98.26 & 82.60 & 98.26 & 79.13 & 66.08 \\ \hline\hline
\textbf{Mean Acc.}  & \textbf{72.47} & 50.92 & 63.57 & 40.92 & 64.95 & 67.57 & 64.11 & 49.40 & 35.46 \\
\textbf{Time (msec)} &5	& 5	& 11	& 6 & 11	& 6	& 5	& 6	& 6
\end{tabular}
}
\end{table}

\begin{table}[!ht]
\begin{center}
\caption{Results using different kernels for the SVM classifier}
\rowcolors{1}{Gray!25}{Gray!5}
\resizebox{\textwidth}{!}{
\begin{tabular}[c]{p{0.7cm}p{0.7cm}p{0.7cm}p{0.9cm}p{0.9cm}p{0.9cm}p{0.9cm}p{0.9cm}p{0.9cm}p{0.9cm}p{0.9cm}}  \hline
\emph{Sub. ID} & \emph{linear} & \emph{RBF} & \emph{chi-square} & \emph{Stand. Euclidean} & \emph{city-block} & \emph{Minko-wski} & \emph{Cheby-shev} & \emph{Cosi-ne} & \emph{Corre-lation} & \emph{Spear-man}\\
 \hline\hline
S001    & 98.55 & \textbf{100.0} & 98.55 & 94.20 & \textbf{100.0} & \textbf{100.0} & 98.55 & \textbf{100.0} & \textbf{100.0} & 92.75 \\
S002    & 87.82 & 78.26 & 83.47 & 89.56 & 73.91 & 67.82 & 46.95 & 75.65 & 74.78 & \textbf{90.43} \\
S003    & 34.78 & 20.28 & 20.28 & 34.78 & 21.73 & 21.73 & 26.08 & 23.18 & 23.18 & \textbf{49.27} \\
S004    & 77.17 & 75.00 & 72.82 & \textbf{81.52} & 76.08 & 69.56 & 71.73 & 77.17 & 78.26 & 79.34 \\
S005    & \textbf{30.43} & 23.47 & 28.69 & 27.82 & 26.08 & 26.95 & 26.95 & 23.47 & 24.34 & 29.56 \\
S006    & 86.08 & 50.43 & 60.00 & 80.00 & 54.78 & 48.69 & 47.82 & 54.78 & 55.65 & \textbf{87.82} \\
S007    & \textbf{60.00} & 70.43 & 73.04 & 54.78 & 71.30 & 70.43 & 71.30 & 73.04 & 72.17 & 52.17 \\
S008    & 31.88 & 33.33 & 40.57 & 39.13 & 37.68 & 34.78 & 20.28 & 34.78 & 36.23 & \textbf{37.68} \\
S009    & \textbf{100.0} & 98.26 & 98.26 & 99.13 & 98.26 & 98.26 & 98.26 & 98.26 & 98.26 & 99.13 \\
S010    & 92.17 & 69.56 & 80.00 & 88.69 & 79.13 & 66.08 & 54.78 & 68.69 & 70.43 & \textbf{93.91} \\
S011    & 98.26 & 97.39 & 97.39 & 98.26 & 97.39 & 96.52 & 93.04 & 96.52 & 96.52 & \textbf{99.13} \\ \hline\hline 
\textbf{Mean Acc.}  & 72.47 & 65.13 & 68.46 & 71.62 & 66.94 & 63.71 & 59.61 & 65.96 & 66.35 & \textbf{73.74} \\
\textbf{Time (msec)} & 5	& 5	& 5	& 5 & 5 & 5 & 5 & 5 & 5 & 5
\end{tabular}
}
\label{tb:comparingKernels}
\end{center}
\end{table}


\subsection{Occipital Channels}
\begin{figure}[!t]
\begin{center}
\includegraphics[width=15.0cm]{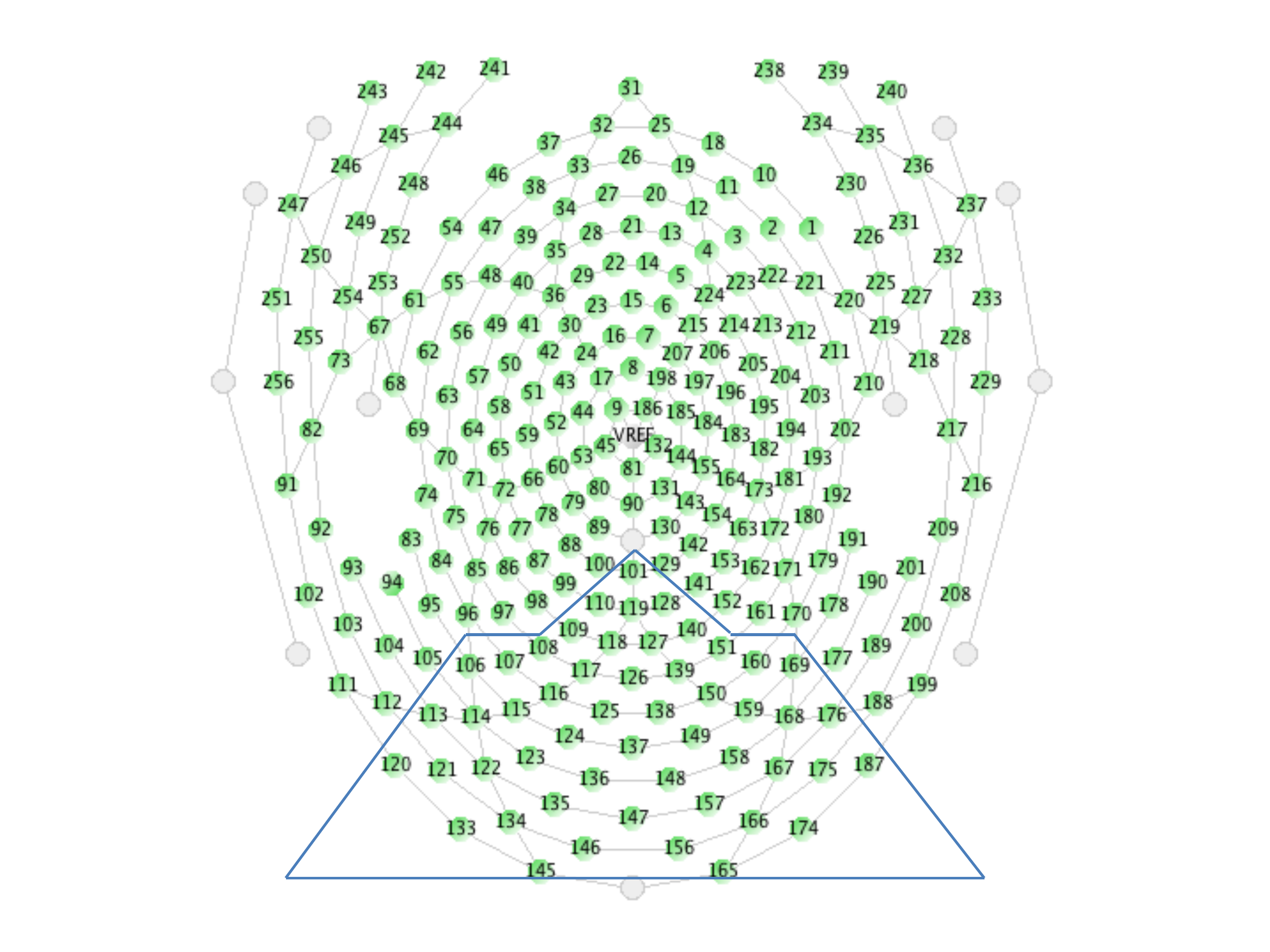}
\caption{Mapping of the 256-channel Sensor Net and identification of the channels lying in the occipital are}
\label{fig:occipitalChannels}
\end{center}
\end{figure}

As previously described, the best locations to acquire SSVEPs are from the occipital lobe. Until 
now all experiments were performed on Oz channel, which is on the midlline of the occipital lobe. 
However, the use of 256-channel Sensor Net for capturing the EEG signal allowed us to investigate 
the classification performance in other parts of the brain that lie in the occipital area. Towards 
this goal, we used the map of the 256-channel Sensor Net depicted in Figure~\ref{fig:occipitalChannels} 
so as to identify the channels that lie in the occipital area. 40 channels were identified, 
which are the ones enclosed by the bounding boxes overlaid on the map of Figure~\ref{fig:occipitalChannels}.  

After identifying the channels lying in the occipital area, the default configuration was 
used to estimate the classification accuracy in each channel. The results are depicted in
Table~\ref{tbl:occipitalChannelsPerformance}. We can see that using electrode O2 results in 
better accuracy compared to Oz and that the use of O1 deteriorates significantly the outcome. 
Furthermore, electrodes close to O2 seem to be more reliable, with channel-138 being the best 
performing electrode with a mean accuracy of 74.42\%, improved approximately by 2.0\% compared
to the baseline configuration. In addition, we should mention that in the extended version of
Table~\ref{tbl:occipitalChannelsPerformance} where the accuracy is estimated per subject, it was 
particularly interesting to observe that the optimal electrode varied per subject with a difference 
that could somehow reach (or exceed) 4\% with respect to the default configuration. Finally, based 
on the results of Table~\ref{tbl:occipitalChannelsPerformance} the EEG signals generated from 
channel 138 (see Figure~\ref{fig:occipitalChannels}) were used in our optimal configuration. 

\begin{table}[!ht]
\begin{center}
\caption{Results using all occipital channels}
\resizebox{\textwidth}{!}{
\rowcolors{1}{Gray!25}{Gray!5}
\begin{tabular}[c]{p{1.25cm}p{1.2cm}p{1.25cm}p{1.2cm}p{1.25cm}p{1.2cm}p{1.25cm}p{1.2cm}} \hline
Electrode & Mean Acc. & Electrode & Mean Acc. & Electrode & Mean Acc.  & Electrode & Mean Acc. \\ \hline
126 (Oz) & 72.47  & 116 (O1)& 66.89 & 133 & 46.40 & 158 & 67.65 \\
138 & 74.42  & 117 & 67.91 & 134 & 56.80 & 159 & 67.97 \\
150 (O2) & 72.49  & 118 & 66.94 & 135 & 63.22 & 165 & 52.47 \\
139 & 73.45  & 119 & 64.88 & 136 & 65.25 & 166 & 61.13 \\
137 & 72.27  & 120 & 44.62 & 145 & 50.88 & 167 & 63.82 \\
149 & 72.74 & 121 & 51.98 & 146 & 58.82 & 168 & 64.89 \\
125 & 71.75 & 122 & 58.07 & 147 & 68.64 & 174 & 49.16 \\
113 & 50.79  & 123 & 60.09 & 148 & 70.69 & 175 & 59.90 \\
114 & 53.88  & 124 & 69.51 & 156 & 64.50 & 176 & 63.73 \\
115 & 60.73  & 127 & 68.67 & 157 & 70.11 & 187 & 54.49 \\
\hline
\end{tabular}
\label{tbl:occipitalChannelsPerformance}
}
\end{center}
\end{table}

\subsection{Optimal configuration}
By combining the best performing algorithms and methods as indicated by the 
comparative evaluations performed in Section~\ref{section:experimentalstudy} 
we can obtain the optimal configuration of Table~\ref{tbl:optimalConfiguration}. 
We can see that there many differences compared to the default configuration of 
Table~\ref{tbl:defaultConfiguration}, such as the incorporation of an IIR-elliptic 
filter instead of an IIR-Chebyshev I, an artifact removal process based on AMUSE, a feature selection algorithm based on SVD, as well as the optimal tuning of PWelch 
for feature extraction and SVM for classification. Finally, in the optimal configuration 
the EEG signals are obtained from channel-138 rather than channel-126 
(see Figure~\ref{fig:occipitalChannels}).

\begin{table}[!ht]
\begin{center}
\caption{Optimal Configuration - Parameter Values}
\rowcolors{1}{Gray!25}{Gray!5}
\begin{tabular}[c]{p{10cm}} \hline\hline
\emph{Channel} \\ 
Channel-138 lying between Oz and  O2 channels (see Figure \ref{fig:occipitalChannels}.) \\ \hline\hline
\emph{Duration} \\ 
Duration of used EEG data: all EEG trials (5sec)\\ \hline\hline
\emph{Filtering} \\ 
Band-pass filter from 5-48Hz using an IIR-elliptic filter 
\\
\hline\hline
\emph{Artifact removal}\\ 
Employment of the AMUSE algorithm for removing the first 15 and the last 4 components before reconstructing the original signal. 
\\ \hline\hline
\emph{Feature extraction}\\ 
PWelch algorithm with nfft=512, segment length=350 and overlap=0.75\\
\hline\hline
\emph{Feature selection} \\
SVD using the 90 features out of 257 of the original feature space. \\
\hline\hline
\emph{Classification} \\ 
SVM classifier using a Kernel based on Spearman correlation and the one-vs-all approach for adopting the binary classifier into our multiclass problem.\\
\hline\hline
\end{tabular}
\label{tbl:optimalConfiguration}
\end{center}
\end{table}

Table~\ref{tbl:performanceOptimalConfiguration} depicts the classification accuracy of 
the optimal configuration compared to the default. We can see that by using the optimal
configuration we manage to introduce an improvement of approximately 7\% 
compared to the default, showing the importance of tuning the employed algorithms in a SSVEP-based BCI. More specifically, with respect to the categorization of the subjects based on their performance (see Section~\ref{subsection:multiparameterselection}) we can see that the main source of this improvement are the subjects coming from the \textit{mid-accurate} category. Indeed, there has been an improvement of $\approx$12\% for S002, $\approx$11\% for S006 and $\approx$30\% for S007. The situation has been also favorable (although less impressive) for the subjects coming from the \textit{poorly-accurate} category with the improvement being $\approx$18\% for S003 and $\approx$6\% for S008, while deteriorating by $\approx$3\% for S005. The subjects coming from the \textit{highly-accurate} category can be considered as stable, with the only exception of S010 that achieved an improvement of $\approx$5\% in the optimal configuration.  

In an attempt to further analyze these findings we may speculate about the following. The poor performance exhibited by the subjects coming from the \textit{mid-accurate} category has been mainly due to the artifacts generated from eye-blinks and other external factors. This was the reason that AMUSE has been particularly effective in improving the performance for two of them (i.e. see S002 and S007 in Table~\ref{tbl:artifactRemoval}). On the other hand, the very low performance of the subjects coming from the \textit{poorly-accurate} category can be attributed to the failure of our system in capturing their EEG signals when reaching a steady-state, either due to problems in impedance (thick hair) or other reasons. In this case, particularly favorable has been the impact of employing the Spearman kernel (see S003 and S008 in Table~\ref{tb:comparingKernels}) indicating that a metric based on correlation can be more appropriate in these situations. Finally, the high performance exhibited by the subjects coming from the \textit{highly-accurate} category proved rather stable across the different experiments. 

Finally, we should note that the improved classification accuracy achieved with the optimal configuration comes at the expense of additional computation cost. More specifically, the optimal configuration needs 70 msecs to reach a decision, while the default configuration needs only 5 msecs. We can see that this is mainly due to the inclusion of the AMUSE method into the signal processing pipeline. However, a total execution time of 70 msecs for the signal processing module of a BCI system can still be considered as an acceptable delay for real-time applications. 


\begin{table}[!ht]
\begin{center}
\caption{Classification accuracy using the optimal configuration.}
\rowcolors{1}{Gray!25}{Gray!5}
\begin{tabular}[c]{ccc}  \hline
\emph{Subject ID} & \emph{Default} & \emph{Optimal} \\ \hline\hline
S001    & \textbf{98.5507}     & 97.1014     \\
S002    & 87.8261     & \textbf{99.1304}     \\
S003    & 34.7826     & \textbf{52.1739}     \\
S004    & 77.1739     & \textbf{78.2609}     \\
S005    & \textbf{30.4348}     & 27.8261     \\
S006    & 86.0870     & \textbf{97.3913}     \\
S007    & 60.0000     & \textbf{89.5652}     \\
S008    & 31.8841     & \textbf{36.2319}     \\
S009    & \textbf{100.000}     & \textbf{100.000}     \\
S010    & 92.1739     & \textbf{97.3913}     \\
S011    & 98.2609     & \textbf{99.1304}     \\ \hline \hline
\textbf{Mean Acc.} & 72.4703 & \textbf{79.4729} \\
\textbf{Time (msec)} &  5 & 70
\end{tabular}
\label{tbl:performanceOptimalConfiguration}
\end{center}
\end{table}
  
\section{Conclusions and opportunities for research}\label{section:conclusions}

It concluding this document it is interesting to highlight some of the areas that we consider 
as particularly favorable for future research based on our experimental findings. First and 
foremost it has been very interesting to observe the variability in the accuracy exhibited by 
the employed algorithms with respect to the examined subjects. More specifically, subjects S003, 
S005 and S008 (all classified in the poorly accurate category, see Section~\ref{subsection:multiparameterselection}) 
were found to be the outliers in the superiority of the best performing algorithms in many different cases. This diversity across subjects has been also observed for the optimal location of the EEG electrodes. All the above, are clear evidence of the potential 
in devising ``calibration'' processes (e.g. running for a small period of time before the actual 
operation) that would allow the BCI system to optimally tune its internal parameters with respect 
to the particularity of the subject.   

Another very interesting future direction that derives from the aforementioned observations is 
the potential of fusing the information coming from different EEG channels in order to devise 
more robust classification schemes. This type of approaches could either take the form of early 
fusion where the EEG signals coming from different channels are combined before delivered to the 
classification scheme, or late fusion where an independent classifier is trained for each EEG 
channel and decisions are taken by jointly considering the output of all classifiers, 
such as in a majority voting scheme.   

Finally, another very important direction for future research has to do with the protocols employed 
for communicating the visual stimuli and the information transfer rate achieved by the system. In our 
experiments, we have only investigated the most simple case where the subject is presented with a single 
box flickering for 5 seconds and the feature extraction algorithm is applied on the full duration of the 
trial. Although this setting allowed us to obtain very clean EEG signals and reach some very concrete 
conclusions about the optimal configuration of a BCI system, it can be considered prohibitive for a 
real-time application. Motivated by this fact our immediate plans for future work include the execution 
of an experimental protocol where more than one visual stimuli are presented simultaneously to the 
subject, as well as the investigation of the trade-off between the classification accuracy and the 
duration of the trial.

\bibliographystyle{ieeetr}
\bibliography{mamem_bib1}

\begin{thebibliography}{10}

\bibitem{acns:2008}
{American Clinical Neurophysiology Society}, ``{Guideline 9B: Recommended
  Standards for Visual Evoked Potentials}.''
  \url{https://www.acns.org/pdf/guidelines/Guideline-9B.pdf}, 2008.

\bibitem{LuckERPbook}
S.~Luck, {\em An Introduction to the Event-Related Potential Technique}.
\newblock MIT Press, 2005.

\bibitem{Guger:2001}
C.~Guger, A.~Schlögl, C.~Neuper, D.~Walterspacher, T.~Strein, and
  G.~Pfurtscheller, ``Rapid prototyping of an eeg-based brain-computer
  interface (bci),'' {\em IEEE Trans. Neural Syst. Rehab. Eng.}, vol.~9, no.~1,
  p.~49–58, 2001.

\bibitem{Pfurtscheller:2006}
G.~Pfurtscheller, R.~Leeb, C.~Keinrath, D.~Friedman, C.~Neuper, C.~Guger, and
  M.~Slater, ``Walking from thought,'' {\em Brain Res.}, vol.~1071, no.~1,
  p.~145–152, 2006.

\bibitem{Hill:2006}
N.~Hill, T.~Lal, M.~Schroder, T.~Hinterberger, B.~Wilhelm, F.~Nijboer,
  U.~Mochty, G.~Widman, C.~Elger, B.~Scholkopf, A.~Kubler, and N.~Birbaumer,
  ``Classifying eeg and ecog signals without subject training for fast bci
  implementation: Comparison of nonparalyzed and completely paralyzed
  subjects,'' {\em IEEE Trans. Neural Syst. Rehab. Eng.}, vol.~14,
  p.~183–186, 2006.

\bibitem{Weiskopf:2004}
N.~Weiskopf, K.~Mathiak, S.~Bock, F.~Scharnowski, R.~Veit, W.~Grodd, R.~Goebel,
  and N.~Birbaumer, ``Principles of a brain-computer interface (bci) based on
  real-time functional magnetic resonance imaging (fmri),'' {\em IEEE Trans.
  Biomed. Eng.}, vol.~51, p.~966–970, 2004.

\bibitem{Yoo:2004}
S.-S. Yoo, T.~Fairneny, N.-K. Chen, S.-E. Choo, L.~Panych, H.~Park, S.-Y. Lee,
  and F.~Jolesz, ``Brain-computer interface using fmri: Spatial navigation by
  thoughts,'' {\em Neuroreport}, vol.~15, no.~10, p.~1591–1595, 2004.

\bibitem{Matthews:2008}
F.~Matthews, B.~Pearlmutter, T.~Ward, C.~Soraghan, and C.~Markham,
  ``Hemodynamics for brain-computer interfaces,'' {\em IEEE Signal Processing
  Magazine}, vol.~25, no.~1, pp.~87--94, 2008.

\bibitem{Schalk:2004}
G.~Schalk, D.~McFarland, T.~Hinterberger, N.~Birbaumer, and J.~Wolpaw,
  ``Bci2000: a general-purpose brain-computer interface (bci) system,'' {\em
  IEEE Transactions on Biomedical Engineering}, vol.~51, pp.~1034--1043, June
  2004.

\bibitem{Amiri:2013}
S.~Amiri, A.~Rabbi, L.~Azinfar, and R.~Fazel-Rezai, ``{A Review of P300, SSVEP,
  and Hybrid P300/SSVEP Brain- Computer Interface Systems},'' in {\em
  Brain-Computer Interface Systems - Recent Progress and Future Prospects}
  (R.~Fazel-Rezai, ed.), ch.~10, InTech, 2013.

\bibitem{Lin2007}
H.-T. Lin, C.-J. Lin, and R.~C. Weng, ``A note on platt's probabilistic outputs
  for support vector machines,'' {\em Machine Learning}, vol.~68, no.~3,
  pp.~267--276, 2007.

\bibitem{Carvalho:2015}
S.~N. Carvalho, T.~B. Costa, L.~F. Uribe, D.~C. Soriano, G.~F. Yared, L.~C.
  Coradine, and R.~Attux, ``Comparative analysis of strategies for feature
  extraction and classification in {SSVEP BCIs},'' {\em Biomedical Signal
  Processing and Control}, vol.~21, pp.~34 -- 42, 2015.

\bibitem{Bakardjian:2010}
H.~Bakardjian, T.~Tanaka, and A.~Cichocki, ``Optimization of \{SSVEP\} brain
  responses with application to eight-command brain computer interface,'' {\em
  Neuroscience Letters}, vol.~469, no.~1, pp.~34--38, 2010.

\bibitem{Chen:2015b}
X.~Chen, Y.~Wang, S.~Gao, T.-P. Jung, and X.~Gao, ``Filter bank canonical
  correlation analysis for implementing a high-speed ssvep-based brain -
  computer interface,'' {\em Journal of Neural Engineering}, vol.~12, no.~4,
  p.~046008, 2015.

\bibitem{Martinez:2007}
P.~Martinez, H.~Bakardjian, and A.~Cichocki, ``Fully online multicommand
  brain-computer interface with visual neurofeedback using ssvep paradigm,''
  {\em Computational intelligence and neuroscience}, vol.~2007, pp.~13--13,
  2007.

\bibitem{Friman:2007}
O.~Friman, I.~Volosyak, and A.~Graser, ``Multiple channel detection of
  steady-state visual evoked potentials for brain-computer interfaces,'' {\em
  IEEE Transactions on Biomedical Engineering}, vol.~54, pp.~742--750, April
  2007.

\bibitem{Molina:2011}
G.~Garcia-Molina and D.~Zhu, ``Optimal spatial filtering for the steady state
  visual evoked potential: Bci application,'' in {\em 5th International
  IEEE/EMBS Conference on Neural Engineering (NER), 2011}, pp.~156--160, April
  2011.

\bibitem{Parini:2009}
S.~Parini, L.~Maggi, A.~Turconi, and G.~Andreoni, ``A robust and self-paced bci
  system based on a four class ssvep paradigm: Algorithms and protocols for a
  high-transfer-rate direct brain communication,'' {\em Intell. Neuroscience},
  pp.~2:1--2:11, Jan 2009.

\bibitem{Wang:2006}
Y.~Wang, R.~Wang, X.~Gao, B.~Hong, and S.~Gao, ``A practical vep-based
  brain-computer interface,'' {\em IEEE Transactions on Neural Systems and
  Rehabilitation Engineering}, vol.~14, pp.~234--240, June 2006.

\bibitem{Diez:2011}
P.~Diez, V.~Mut, E.~A. Perona, and E.~L. Leber, ``Asynchronous bci control
  using high-frequency ssvep,'' 2011.

\bibitem{Vilic:2013}
A.~Vilic, T.~Kjaer., C.~Thomsen., S.~Puthusserypady, and H.~Sorensen, ``Dtu bci
  speller: An ssvep-based spelling system with dictionary support,'' in {\em
  35th Annual International Conference of the IEEE Engineering in Medicine and
  Biology Society (EMBC), 2013}, pp.~2212--2215, 2013.

\bibitem{Muller:2005}
G.~R. Muller-Putz, R.~Scherer, C.~Brauneis, and G.~Pfurtscheller,
  ``Steady-state visual evoked potential (ssvep)-based communication: impact of
  harmonic frequency components,'' {\em Journal of Neural Engineering}, vol.~2,
  no.~4, pp.~123--130, 2005.

\bibitem{Jia:2011}
C.~Jia, X.~Gao, B.~Hong, and S.~Gao, ``Frequency and phase mixed coding in
  ssvep-based brain--computer interface,'' {\em Biomedical Engineering, IEEE
  Transactions on}, vol.~58, pp.~200--206, Jan 2011.

\bibitem{Guger:2012}
C.~Guger, B.~Allison, B.~Grosswindhager, R.~R.~Pruckl, C.~Hintermuller,
  C.~Kapeller, M.~Bruckner, G.~Krausz, and G.~Edlinger, ``How many people could
  use an ssvep bci?,'' {\em Frontiers in Neuroscience}, vol.~6, no.~169, 2012.

\bibitem{Resalat:2013}
S.~N. Resalat and S.~K. Setarehdan, ``An improved ssvep based bci system using
  frequency domain feature classification,''

\bibitem{Rajesh:2014}
S.~Rajesh and B.~Haseena, ``Comparison of ssvep signal classification
  techniques using svm and ann models for bci applications,'' {\em
  International Journal of Information and Electronics Engineering}, vol.~4,
  January 2014.

\bibitem{Zhang:2013}
Y.~Zhang, G.~Zhou, J.~Jin, M.~Wang, X.~Wang, and A.~Cichocki, ``L1-regularized
  multiway canonical correlation analysis for ssvep-based bci,'' {\em IEEE
  Transactions on Neural Systems and Rehabilitation Engineering}, vol.~21,
  pp.~887--896, Nov 2013.

\bibitem{Zhang:2012b}
Y.~Zhang, J.~Jin, X.~Qing, B.~Wang, and X.~Wang, ``Lasso based stimulus
  frequency recognition model for ssvep bcis,'' {\em Biomedical Signal
  Processing and Control}, vol.~7, no.~2, pp.~104 -- 111, 2012.

\bibitem{Lotte:2007}
F.~Lotte, M.~Congedo, A.~Lecuyer, F.~Lamarche, and B.~Arnaldi, ``A review of
  classification algorithms for eeg-based brain - computer interfaces,'' {\em
  Journal of Neural Engineering}, vol.~4, no.~2, p.~R1, 2007.

\bibitem{OppenheimDSP:1999}
A.~V. Oppenheim, R.~Schafer, and J.~Buck, {\em Discrete-time Signal Processing
  (2Nd Ed.)}.
\newblock Prentice-Hall, Inc., 1999.

\bibitem{ProakisDSP:1996}
J.~Proakis and D.~Manolakis, {\em Digital Signal Processing (3rd Ed.):
  Principles, Algorithms, and Applications}.
\newblock Prentice-Hall, Inc., 1996.

\bibitem{White:2000}
S.~White, {\em Digital Signal Processing: A Filtering Approach}.
\newblock Delmar Cengage Learning, 2000.

\bibitem{Choi:2005}
S.~Choi, A.~Cichocki, H.-M. Park, and S.-Y. Lee, ``{Blind Source Separation and
  Independent Component Analysis: A Review},'' {\em Neural Information
  Processing - Letters and Reviews}, vol.~6, pp.~1--57, jan 2005.

\bibitem{Hyvarinen:2000}
A.~Hyvarinen and E.~Oja, ``Independent component analysis: Algorithms and
  applications,'' {\em Neural Networks}, vol.~13, pp.~411--430, 2000.

\bibitem{Jung:2001a}
T.~Jung, S.~Makeig, M.~Mckeown, A.~Bell, T.~Lee, and T.~Sejnowski, ``{Imaging
  brain dynamics using independent component analysis},'' {\em Proc. IEEE},
  vol.~89, no.~7, 2001.

\bibitem{Jung:2001b}
T.~Jung, S.~Makeig, M.~Westerfield, J.~Townsend, E.~Courchesne, and
  T.~Sejnowski, ``Analysis and visualization of single-trial event-related
  potentials,'' {\em Human Brain Mapping}, vol.~14, pp.~166--185, 2001.

\bibitem{stoica2005spectral}
P.~Stoica and R.~Moses, {\em Spectral Analysis of Signals}.
\newblock Pearson Prentice Hall, 2005.

\bibitem{MallatWTS:2008}
S.~Mallat, {\em A Wavelet Tour of Signal Processing, Third Edition: The Sparse
  Way}.
\newblock Academic Press, 3rd~ed., 2008.

\bibitem{Shannon:1948}
C.~Channon, ``A mathematical theory of communication,'' {\em The Bell System
  Technical Journal}, vol.~27, no.~7, pp.~379 -- 423.

\bibitem{Brown:2012}
G.~Brown, A.~Pocock, M.-J. Zhao, and M.~Luj\'{a}n, ``{Conditional Likelihood
  Maximisation: A Unifying Framework for Information Theoretic Feature
  Selection},'' {\em J. Mach. Learn. Res.}, vol.~13, pp.~27--66, Mar. 2012.

\bibitem{Platt99}
J.~C. Platt, ``Probabilistic outputs for support vector machines and
  comparisons to regularized likelihood methods,'' in {\em Advances in Large
  Margin Classifiers}, pp.~61--74, MIT Press, 1999.

\bibitem{eeg:300}
``{Geodesic EEG System 300}.''
  \url{https://www.egi.com/clinical-division/clinical-division-clinical-products/ges-300}.

\bibitem{stimtracker}
``{Cedrus StimTracker}.'' \url{http://cedrus.com/stimtracker/}.

\bibitem{Georgiadis2016}
K.~Georgiadis, G.~Liaros, V.~P. Oikonomou, E.~Chatzilari, K.~Adam,
  S.~Nikolopoulos, and I.~Kompatsiaris, ``Mamem eeg ssvep dataset i (256
  channels, 11 subjects, 5 frequencies).''
  \url{https://dx.doi.org/10.6084/m9.figshare.2068677.v1}, 2016.

\bibitem{Liaros2016}
G.~Liaros, V.~P. Oikonomou, K.~Georgiadis, E.~Chatzilari, K.~Adam,
  S.~Nikolopoulos, and I.~Kompatsiaris, ``ssvep-eeg-processing-toolbox.''
  \url{https://github.com/MAMEM/ssvep-eeg-processing-toolbox}, 2016.

\bibitem{CC01a}
C.-C. Chang and C.-J. Lin, ``{LIBSVM}: A library for support vector machines,''
  {\em ACM Transactions on Intelligent Systems and Technology}, vol.~2,
  pp.~27:1--27:27, 2011.
\newblock Software available at \url{http://www.csie.ntu.edu.tw/~cjlin/libsvm}.

\bibitem{Mussel:2012}
C.~Müssel, L.~Lausser, M.~Maucher, and H.~Kestler, ``Multi-objective parameter
  selection for classifiers,'' {\em Journal of Statistical Software}, vol.~46,
  no.~1, 2012.

\bibitem{Arlot:2010}
S.~Arlot and A.~Celisse, ``A survey of cross-validation procedures for model
  selection,'' {\em Statistics Surveys}, vol.~4, pp.~40--79, 2010.

\bibitem{Rice:1991}
J.~A. Rice and B.~W. Silverman, ``Estimating the mean and covariance structure
  nonparametrically when the data are curves,'' {\em Journal of the Royal
  Statistical Society. Series B (Methodological)}, vol.~53, no.~1,
  pp.~233--243, 1991.

\bibitem{Xu:2012}
G.~G.~Xu and J.~Huang, ``Asymptotic optimality and efficient computation of the
  leave-subject-out cross-validation,'' {\em The Annals of Statistics},
  vol.~40, pp.~3003--3030, 12 2012.

\bibitem{Fazli:2009}
S.~Fazli, F.~Popescu, M.~Danóczy, B.~Blankertz, K.~Müller, and C.~Grozea,
  ``Subject-independent mental state classification in single trials,'' {\em
  Neural Networks}, vol.~22, pp.~1305--1312, 11 2009.

\end{thebibliography}

\end{document}